\documentclass[sigplan,twocolumn]{acmart}
\renewcommand\footnotetextcopyrightpermission[1]{}
\usepackage{algorithm}
\usepackage{algpseudocode}
\usepackage{booktabs}
\usepackage{arydshln}
\usepackage{graphicx}
\usepackage{subfig}
\usepackage{svg}
\usepackage{todonotes}
\usepackage{textcomp}
\usepackage{xcolor}
\usepackage{xspace}
\usepackage{csquotes}
\usepackage[htt]{hyphenat}
\usepackage{tablefootnote}
\definecolor{darkgreen}{rgb}{0,0.75,0}
\definecolor{darkblue}{rgb}{0,0,0.75}
\definecolor{darkorange}{rgb}{0.8,.33,0}

\newcommand{\etal}{\xspace\textit{et al.}\xspace}
\newcommand{\dynasplit}{\texttt{DynaSplit}\xspace}
\newcommand{\solver}{\emph{\dynasplit Solver}\xspace}
\newcommand{\controller}{\emph{\dynasplit Controller}\xspace}

\usepackage{tcolorbox}
\usepackage[inline]{enumitem}
\newtcbox{\hl}[1][]{on line, colback=yellow!50, colframe=yellow!50, boxrule=0pt, sharp corners, boxsep=0pt, left=1pt, right=1pt, top=0pt, bottom=0pt, #1}

\AtBeginDocument{%
  }

\begin{document}

\title{
DynaSplit: A Hardware-Software Co-Design Framework for Energy-Aware Inference on Edge
}
\author{Daniel May}
 \email{daniel.may@tuwien.ac.at}
\orcid{0009-0009-8434-4639}
\affiliation{%
   \institution{Institute of Information Systems Engineering, TU Wien}
   \city{Vienna}
   \country{Austria}
 }

 \author{Alessandro Tundo}
 \email{alessandro.tundo@tuwien.ac.at}
 \orcid{0000-0001-8840-8948}
 \affiliation{%
   \institution{Institute of Information Systems Engineering, TU Wien}
   \city{Vienna}
   \country{Austria}
 }

 \author{Shashikant Ilager}
 \email{s.s.ilager@uva.nl}
 \orcid{0000-0003-1178-6582}
 \affiliation{%
   \institution{Informatics Institute, University of Amsterdam}
   \city{Amsterdam}
   \country{Netherlands}
 }

 \author{Ivona Brandic}
 \email{ivona.brandic@tuwien.ac.at}
 \orcid{0000-0001-7424-0208}
 \affiliation{%
   \institution{Institute of Information Systems Engineering, TU Wien}
   \city{Vienna}
   \country{Austria}
 }

\renewcommand{\shortauthors}{May et al.}

\begin{abstract}
The deployment of ML models on edge devices is challenged by limited computational resources and energy availability. While split computing enables the decomposition of large neural networks (NNs) and allows partial computation on both edge and cloud devices, identifying the most suitable split layer and hardware configurations is a non-trivial task. This process is in fact hindered by the large configuration space, the non-linear dependencies between software and hardware parameters, the heterogeneous hardware and energy characteristics, and the dynamic workload conditions.
To overcome this challenge, we propose \dynasplit, a two-phase framework that dynamically configures parameters across both software (i.e., split layer) and hardware (e.g., accelerator usage, CPU frequency). During the \textit{Offline Phase}, we solve a multi-objective optimization problem with a meta-heuristic approach to discover optimal settings. During the \textit{Online Phase}, a scheduling algorithm identifies the most suitable settings for an incoming inference request and configures the system accordingly.
We evaluate \dynasplit using popular pre-trained NNs on a real-world testbed. Experimental results show a reduction in energy consumption up to 72\% compared to cloud-only computation, while meeting $\sim90\%$ of user request's latency threshold compared to baselines.

\end{abstract}



\maketitle
\section{Introduction}\label{sec:introduction}

Edge computing provides resources at the network edge for latency and privacy-sensitive applications, often relying on Machine Learning (ML) models for prediction and analytics~\cite{edge-ai-book, DBLP:conf/eurosys/LujicMPBLB21, DBLP:journals/tie/LinLW19, DBLP:conf/icmla/SchwartzSWP21}. This paradigm, also known as Edge AI~\cite{edge-ai-book, DBLP:journals/iotj/DengZFYDZ20, DBLP:journals/ccr/DingPMABDHKLMMO22} enables smart applications on the edge. However, model inference is resource and power-intensive, while edge nodes are often constrained and operate in unreliable power environments~\cite{edge-ai-book, DBLP:journals/ton/ChenJLF16, DBLP:journals/sensors/CallebautLMOSP21}. Therefore, it is crucial to develop energy-efficient inference techniques for Edge AI.

Edge applications are typically decomposed into multiple services and deployed across end-user devices, edge nodes, and remote clouds. This distributed, microservice-based approach allows latency-sensitive processing on edge nodes while utilizing cloud resources for scalability~\cite{DBLP:journals/tmc/WangLLLLH20,TULI201922,FORTI2021605}. ML models, particularly neural networks (NNs), can similarly be split into sequential tasks or layers, with computation distributed across edge and cloud resources based on workload and availability conditions.

However, partitioning NN-based models for Edge AI inference presents significant challenges due to hardware intricacies, model complexity, and application latency requirements. \emph{First}, a nonlinear relationship exists between split points and resulting latency and energy consumption~\cite{KangHGRMMT17}. \emph{Second}, the model architecture constrains possible split points, especially in complex state-of-the-art models with Direct Acyclic Graph (DAG) structures~\cite{DBLP:conf/infocom/HuBWL19, DBLP:journals/imwut/ZhangLLGWWDW20, DBLP:conf/kdd/Banitalebi-Dehkordi21}. \emph{Third}, hardware characteristics (i.e., CPU/GPU frequency) and network conditions significantly impact latency and energy consumption. \emph{Finally}, varying Quality of Service (QoS) requirements create dynamic workload levels. Therefore, identifying dynamic split points and opportunistically configuring parameters across hardware and software domain 
is crucial for energy-efficient settings that satisfy application QoS.

Existing works on energy-aware inference in edge computing focus on model compression techniques, such as pruning, quantization, and knowledge distillation~\cite{DBLP:journals/tnn/WangLGZ24, DBLP:conf/eccv/HeLLWLH18, DBLP:conf/eccv/YangHCZGSSA18, DBLP:conf/fpga/DingWLXW019, DBLP:conf/dcc/GengFZLAPC19}. However, these approaches affect model accuracy~\cite{DBLP:journals/air/ChoudharyMGS20, DBLP:journals/pieee/DengLHSX20, 10.1145/3552326.3587459} and require running entire models on edge nodes. Studies exploring NN splitting between mobile devices and cloud~\cite{KangHGRMMT17, EshratifarAP21, PagliariCMP21, DBLP:conf/sensys/Yao0LWLSA20, DBLP:conf/wowmom/MatsubaraCSLR22} lack dynamic adaptability~\cite{DBLP:conf/islped/EshratifarEP19, DBLP:conf/mobicom/MatsubaraBCL019, DBLP:journals/imwut/ZhangLLGWWDW20} and an integrated hardware-software parameter configuration. Current research also overlooks state-of-the-art models like transformer-based networks and neglects cloud-side energy usage, focusing solely on edge device consumption.

\emph{To address these challenges, we propose \dynasplit, an energy-aware inference framework for Edge AI using a hardware-software co-design method.} First, we identify parameters across hardware and software domains affecting edge resource model performance and energy consumption. Then, we propose an optimization method to determine a suitable configuration for run-time inference, meeting QoS requirements while ensuring accuracy and reducing energy consumption.

\dynasplit operates in two phases,  offline and online. In offline, we formulate a Multi-Objective Optimization Problem (MOOP), which is solved leveraging a meta-heuristic algorithm to obtain suitable configurations for each employed ML model. Online, a scheduling algorithm dynamically configures the system using such configurations to execute user inference requests. Our approach identifies configurations across hardware and software domains, including optimal layer-wise split points. At the hardware level, we employ Dynamic Voltage Frequency Scaling (DVFS) to fine-tune CPU and accelerator performance, controlling execution speed and power consumption.

We implemented \dynasplit on a prototype testbed with a Raspberry Pi 4B, Google Coral Edge TPU, and a remote GPU-equipped node belonging to a research cloud testbed. In our testbed experiments, we used state-of-the-art NNs (i.e., VGG16 and ViT) and conducted larger scalability tests in simulation.
Results show that \dynasplit can reduce energy consumption up to 72\% when compared to cloud-only computation, while meeting on average $\sim90\%$ of the request latency thresholds.
\emph{Our empirical study provides the following key findings}: (i) smaller models optimized for mobile devices do not benefit from split computing; (ii) larger models benefit from split computing, but the hardware-level configuration significantly affects the inference performance; (iii) identifying the optimal NN split point is challenging due to difficulties in estimating the layer-specific run-time performance (i.e., due to significant variation in intermediate output sizes and layer optimizations).

The paper is organized as follows: Section~\ref{sec:motivation} presents the motivational scenario and initial observations. Section~\ref{sec:problem-formulation} outlines the problem statement. Section~\ref{sec:approach} describes \dynasplit's system model, methodology, and scheduling algorithm. Section~\ref{sec:implementation} details the implementation. Section~\ref{sec:empirical-evaluation} discusses the experimental setup and results. Section~\ref{sec:related-work} reviews related work, and Section~\ref{sec:conclusions} concludes with future directions.

\section{Motivation and Preliminary Study}\label{sec:motivation}

\subsection{Motivational Scenario}

\begin{figure}[!ht]
    \centering
    \includegraphics[width=\linewidth]{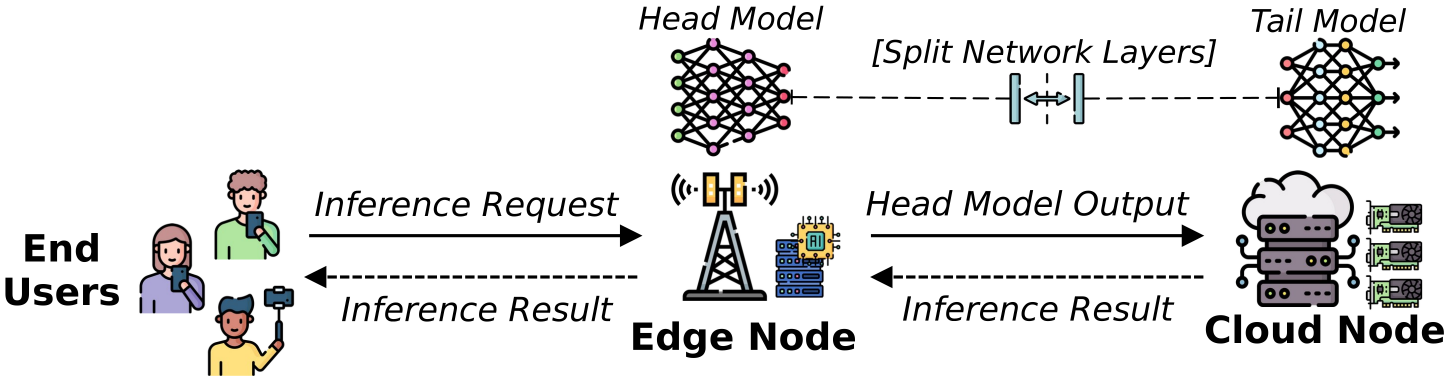}
    \caption{The illustration depicts our motivational scenario in which users send inference requests to an edge application capable of splitting inference between edge and cloud.%
    }
    \label{fig:motivational-scenario}
\end{figure}

Figure~\ref{fig:motivational-scenario} illustrates our motivational scenario where end users employ an edge application for image classification tasks. The application uses an NN and can opportunistically utilize cloud resources. At runtime, it determines whether to execute fully on the edge node (edge-only), fully on a cloud node (cloud-only), or split the NN between edge and cloud. 

Contrary to expectations, edge-only execution does not always reduce inference time compared to cloud-only execution. Both edge-only and cloud-only deployments have shown limitations in inference time and energy consumption~\cite{EshratifarAP21, KangHGRMMT17, PagliariCMP21}. Application performance depends on multiple runtime configurations influencing QoS and energy consumption~\cite{DBLP:conf/kbse/TundoMIBBM23, DBLP:journals/adhoc/HorcasPF19}. These configurable parameters include both hardware (e.g., CPU frequency, using hardware accelerator) and software factors (e.g., model architecture, NN split point). The non-linear relationships and complex behaviors among parameters presents significant challenges~\cite{Hanafy2021, DBLP:journals/tc/Nunez-Yanez19}.

\subsection{Preliminary Observations and Insights}

\begin{figure*}[!ht]
  \centering
  \subfloat[]{\includegraphics[width=0.2\textwidth]{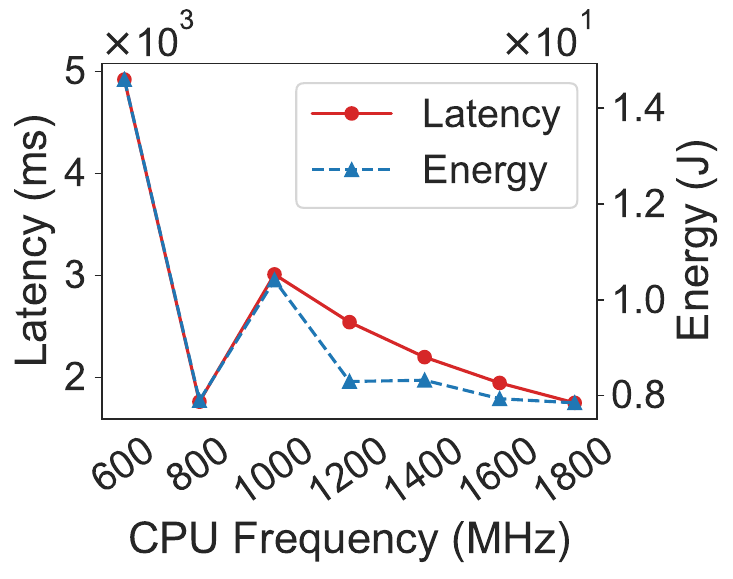}\label{fig:cpu_effects}}
  \subfloat[]{\includegraphics[width=0.2\textwidth]{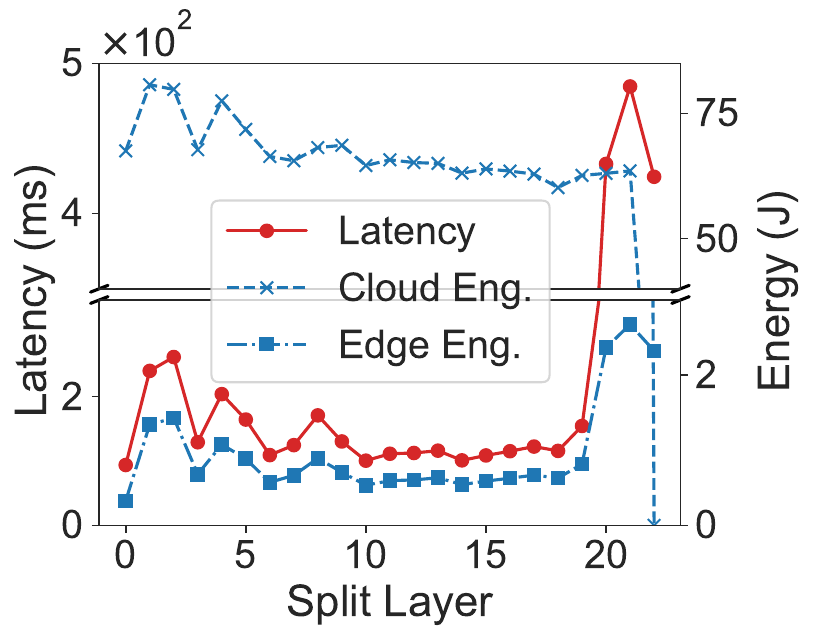}\label{fig:split_layer_effects}}
  \subfloat[]{\includegraphics[width=0.2\textwidth]{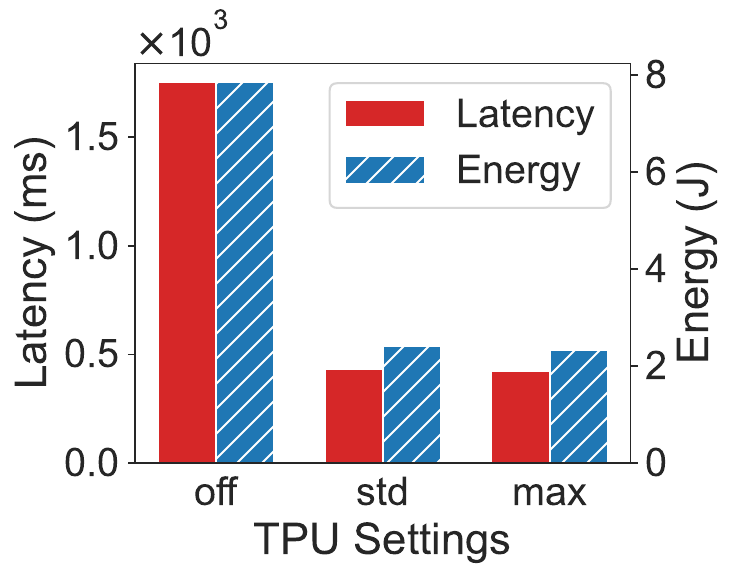}\label{fig:tpu_settings}}
  \subfloat[]{\includegraphics[width=0.2\textwidth]{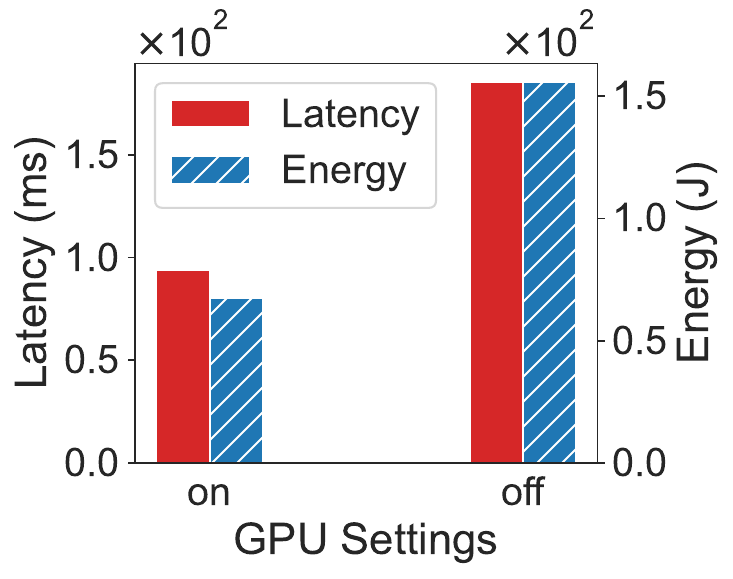}\label{fig:gpu_settings}}
  \subfloat[]{\includegraphics[width=0.2\textwidth]{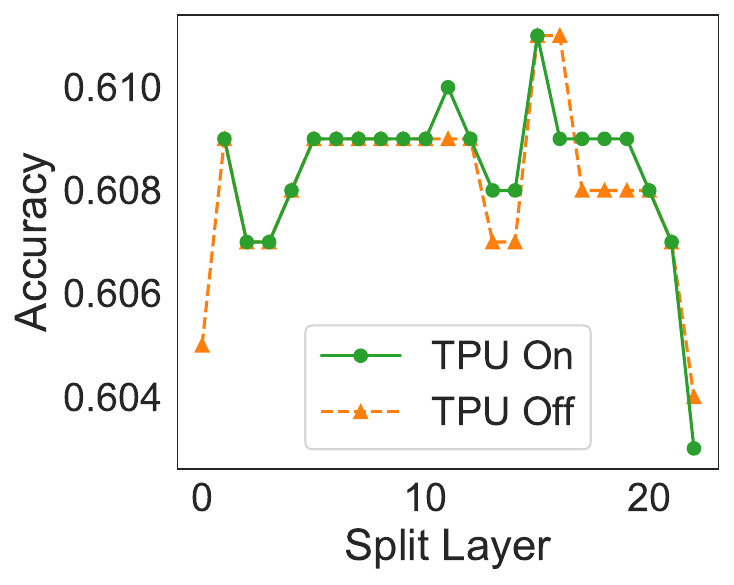}\label{fig:accuracy_effects}}
  \caption{Impact of different configuration parameters on inference latency, energy consumption, and accuracy for the VGG16 network~\cite{vgg}. The reported results are  averaged over 1,000 inferences.}
  \label{fig:configuration_effects}
\end{figure*}
We studied several NNs in split computing environments configuring different hardware and software parameters to further understand the aforementioned scenario with empirical results. Accordingly, we set up an edge-cloud testbed comprising a Raspberry Pi 4B (quad-core ARMv8, 8GB of RAM), a Google Coral TPU as an edge accelerator, a cloud node (2x Intel Xeon E5-2698 v4 CPUs, 512 GiB RAM) equipped with 8 NVIDIA Tesla V100 GPUs (with only one used in practice). We tested four different pre-trained NNs on the ImageNet dataset~\cite{ILSVRC15}, including ResNet50~\cite{resnet} (0.85 million parameters), MobileNetV2~\cite{mobilenetv2}, VGG16~\cite{vgg} with 138 million parameters, and Vision Transformer (ViT)~\cite{vit} with 86 million parameters.  We produced 1,000 inference requests in total as a user workload by randomly selecting images from the ImageNet validation dataset. Furthermore, we collected metrics on  latency, energy consumption, and accuracy.

In our initial experiments, we found that split computing is particularly effective for NNs with a large number of parameters (i.e., VGG16 and ViT). Smaller models (i.e., ResNet50 and MobileNetV2) did not exhibit any benefit from split computing. Smaller models execute faster and consume less power in edge-only deployments. However,  VGG16 and ViT, demonstrated substantial improvements in latency when utilizing both edge and cloud resources, highlighting the potential of split computing for larger models. This aligns with our motivation to explore runtime configurations that leverage both edge and cloud capabilities, enhancing the performance of computationally intensive models while reducing overall energy consumption. Based on these findings, we choose VGG16 and ViT as candidate networks for this study.

Furthermore, we use the VGG16  as an exemplary model to perform an in-depth analysis of the effect of various parameters on inference tasks. VGG16 is a Convolutional Neural Network (CNN) comprising up to 22 layers in its Keras open-source implementation. We consider that the following runtime configurations are possible: (i) adjusting the CPU frequency of the edge node, (ii) determining whether to utilize a hardware accelerator (i.e., a TPU) on the edge node and (iii) dividing the CNN architecture into the dynamic head and tail model, allowing for the execution of certain layers on the edge node and the remainder on a cloud node.

Figure~\ref{fig:configuration_effects} shows the impact of specific configuration parameters on the average latency (in milliseconds), the average energy consumption (in Joules), and the average precision of the inference requests.

Figure~\ref{fig:cpu_effects} shows the results when executed in an edge-only mode by configuring the Raspberry Pi 4B's CPU frequency, without using a TPU edge accelerator. As observed, increasing the CPU frequency leads to a reduction in both latency and energy consumption. The energy consumption initially decreases significantly, but the reduction becomes less pronounced as the CPU frequency increases further.  We observed some outliers at 800 MHz despite multiple runs, which do not follow the general trend.

The effect of the model's split layer on latency and energy consumption can be seen in Figure~\ref{fig:split_layer_effects}. For example, splitting the model at layer 5 results in the execution of the first five layers on the edge node, while the remaining are executed on the cloud node. In this scenario, the TPU is used as an edge accelerator with the maximum possible frequency (500MHz), whereas the CPU frequency is set to 1,800 MHz. The cloud also uses the GPU as an accelerator. As observed, both latency and energy consumption exhibit a similar trend. However, it is evident that selecting an appropriate split point is challenging, as latency and energy consumption are not directly related to the split point.

We also analyzed the effect of the edge acceleration in edge-only settings. Here, the TPU is either off, operating at the standard frequency of 250 MHz (std), or  at the maximum frequency of 500 MHz (max). The effect of these configurations on energy and latency  can be observed in Figure~\ref{fig:tpu_settings}.
Although the TPU draws more power, we can observe that the average energy consumption is $\sim3\times$ lower (compared to the computation on the CPU) due to the accelerated computation and the resulting reduction in the average inference time. There  are  also no significant differences between 250MHz (std) and 500MHz (max) TPU configurations for this particular network. Similarly, we study the effect of cloud acceleration with and without GPU in cloud-only deployment (Figure~\ref{fig:gpu_settings}). As observed,  GPU acceleration  significantly decreases both latency and cloud energy consumption.

Finally, we investigate the effect of edge acceleration and split layer on inference accuracy. There is expected to be a drop in accuracy with an increasing number of layers executed on the edge node because the model has been quantized (8-bit integer in our case) in order to facilitate its execution on a resource-constrained edge device~\cite{DBLP:journals/pieee/DengLHSX20, DBLP:journals/air/ChoudharyMGS20}. Our results show negligible accuracy variations, all within the sub-percent range as observed in Figure~\ref{fig:accuracy_effects}. There is a slight accuracy drop when the last layers are run on the edge, but no clear pattern emerges between using the TPU or CPU, with minimal differences likely due to numerical effects.

Our observations indicate that a combination of integrated hardware-software parameters highly influences inference tasks' latency and energy consumption while it has a negligible effect on accuracy. This highlights the need for advanced runtime configuration and scheduling techniques for NN-based inference in edge-cloud environments that must meet QoS requirements while optimizing energy consumption.

\section{Definitions and Problem Formulation}\label{sec:problem-formulation}
This section presents the essential components of our system model and defines the underlying optimization problem.

\subsection{Model Partitioning}
Consider a NN $M$ consisting of $L$ layers (excluding the input and output layers), which can be partitioned into two segments: the head segment $M_h$ (the first $k$ layers) and the tail segment $M_t$ (the remaining $L - k$ layers). The head segment $M_h$ can be executed on the edge node, while the tail segment $M_t$ can be executed on the cloud node. The edge node transfers the intermediate output vector from head segment $M_h$ to a cloud node, for further computations through tail segment $M_t$. Here, the split layer $k$ can take any value between 0 and $L$, with the following special cases:
\begin{enumerate*}[label=(\roman*)]
    \item $k = 0$: this represents the cloud-only case, where the entire inference is executed in the cloud, and the edge node transfers the input data;
    \item $k = L$: this represents the edge-only case, where the entire inference is performed on the edge node, and no data is transferred to the cloud.
\end{enumerate*}

The ability to adjust $k$ enables the dynamic partitioning of the NN inference process between the edge and the cloud, playing a critical role in balancing latency, energy consumption, and accuracy based on the constraints of the model and system, as well as the application's QoS requirements. 

Here, we assume that edge nodes have limited computational resources compared to cloud nodes, resulting in limited capacity to support a high number of latency-sensitive user requests.     
Similarly, cloud resources provide elasticity, where resources can dynamically scale, and energy is primarily consumed during active computation, not during data transmission or idle periods. New cloud application deployment models, such as serverless computing, enable such on-demand computation services ~\cite{serverless_inference}.

\subsection{Configuration Space}

Table~\ref{tab:search-space} shows the specific configurations across the hardware and software domains we used in this study. Our objective is to identify the most suitable configuration value for a given inference task across these parameters. Therefore, our total configuration space $X$ for this problem instance includes relevant software and hardware parameters such as split layer, CPU frequency settings for the edge, edge hardware accelerator usage, and cloud GPU usage. 

\begin{table}[!ht]
\centering
\caption{The type and domain of the hardware and software parameters.}
\label{tab:search-space}
\resizebox{\linewidth}{!}{%
\begin{tabular}{@{}lll@{}}
\toprule
\textbf{Parameter}                        & \textbf{Parameter Type}                                                              & \textbf{Domain}                     \\ \midrule
CPU Frequency ($CPU_f$)                     & \begin{tabular}[c]{@{}l@{}}Numerical\\ (low: 0.6, high: 1.8, step: 0.2)\end{tabular} & \{0.6, 0.8, 1.0, 1.2, 1.4, 1.6, 1.8\} \\ \midrule
Edge TPU Frequency ($TPU_f$)                       & Categorical                                                                          & \{off, std, max\}                          \\ \midrule
Use Cloud GPU (GPU)                             & Categorical                                                                          & \{Yes, No\}                           \\ \midrule
VGG16~\cite{vgg} Split Layer ($L_{VGG}$)              & \begin{tabular}[c]{@{}l@{}}Numerical\\  (low: 0, high: 22, step: 1)\end{tabular}      & \{0, 1, 2, \dots, 22\}                \\ \hdashline
Vision Transformer~\cite{vit} Split Layer ($L_{ViT}$) & \begin{tabular}[c]{@{}l@{}}Numerical\\ (low: 0, high: 19, step: 1)\end{tabular}      & \{0, 1, 2, \dots, 19\}                \\ \bottomrule
\end{tabular}%
}
\end{table}\textbf{}

\subsection{Latency Model}

The total inference time $T_{\text{inf}}(x)$ for a given configuration $x \in X$  is the sum of the $T_{\text{edge}}(x)$ (edge latency), 
$T_{\text{net}}(x)$ (network latency), and $T_{\text{cloud}}(x)$ (cloud latency). Here, 
$T_{\text{edge}}(x)$ is the time required to execute the head segment $M_h$ on the edge node.
$T_{\text{net}}(x)$ is the time required to transfer data from the edge node to a cloud node and  to receive the results. Finally, $T_{\text{cloud}}(x)$ is the time required to execute the tail segment $M_t$ on cloud node. Thus, the total inference latency is:
\[
T_{\text{inf}}(x) = T_{\text{edge}}(x) + T_{\text{net}}(x) + T_{\text{cloud}}(x)
\]
With the following special cases:
\begin{enumerate*}[label=(\roman*)]
    \item when $k = 0$ (Cloud-Only Inference): $T_{\text{edge}}(x)$ is reduced, as the result of the NN is fully computed in the cloud. However, minimal processing, such as preparing data for transfer, still takes place on the edge;
    \item when $k = L$ (Edge-Only Inference): $T_{\text{cloud}}(x) = 0$ and $T_{\text{net}}(x) = 0$, since computation of all layers is performed at the edge node. No data transfer to or from the cloud is needed. 
\end{enumerate*}

\subsection{Energy Model}
The total energy consumption $E_{\text{inf}}(x)$ consists of the energy consumed by both the edge and cloud computation. 
For the edge node, we compute the energy over the entire inference duration, starting at $t_0$ (time at Layer 0) and ending at $t_{\text{inf}}$, corresponding to the total duration of the inference.
For the cloud node, we compute the energy only during the active computation phase, (i.e., between $t_{\text{net1}}$, the start of the tail segment in the $k^{th}$ layer) and $t_{\text{net2}}$, the end of the tail segment computation at the $L^{th}$ layer, respectively.

Although we account for the energy consumed by the edge and cloud devices during their respective computation phases, we do not include the energy consumed by network components (e.g., routers and switches). This cost is not considered because it is either negligible or difficult to measure in practice ~\cite{6819864}. This exclusion reflects practical challenges in measuring the energy of the network components, which is considerably low compared to the compute energy.

Thus, the total energy consumption is given by:
\[
E_{\text{inf}}(x) = 
\begin{cases} 
\int_{t_0}^{t_{\text{inf}}} P_{\text{edge}}(t, x) \, dt &  \hspace{-1cm} \text{if edge-only}, \\
\int_{t_0}^{t_{\text{inf}}} P_{\text{edge}}(t, x) \, dt + \int_{t_{\text{net1}}}^{t_{\text{net2}}} P_{\text{cloud}}(t, x) \, dt & \text{else}.
\end{cases}
\]

where:
\begin{enumerate*}[label=(\roman*)]
    \item $t_0$ is the time when the overall inference process begins (starting edge inference or preparing data to send to the cloud);
    \item $t_{\text{net1}}$ is the point in time when all data has been transferred from the edge to the cloud, and the cloud is ready to start processing. This point only exists if the cloud component is involved, i.e., $k < L$;
    \item $t_{\text{net2}}$ is the time when the cloud finishes processing and sends the results back to the edge. Like $t_{\text{net1}}$, this point only exists if the cloud component is involved ($k < L$);
    \item $t_{\text{inf}}$ is the time when the inference process ends, either when the edge node has computed all results (in the edge-only case) or when the results have been received from the cloud node.
\end{enumerate*}

\begin{figure*}[!ht]
    \centering
    \includegraphics[width=\linewidth]{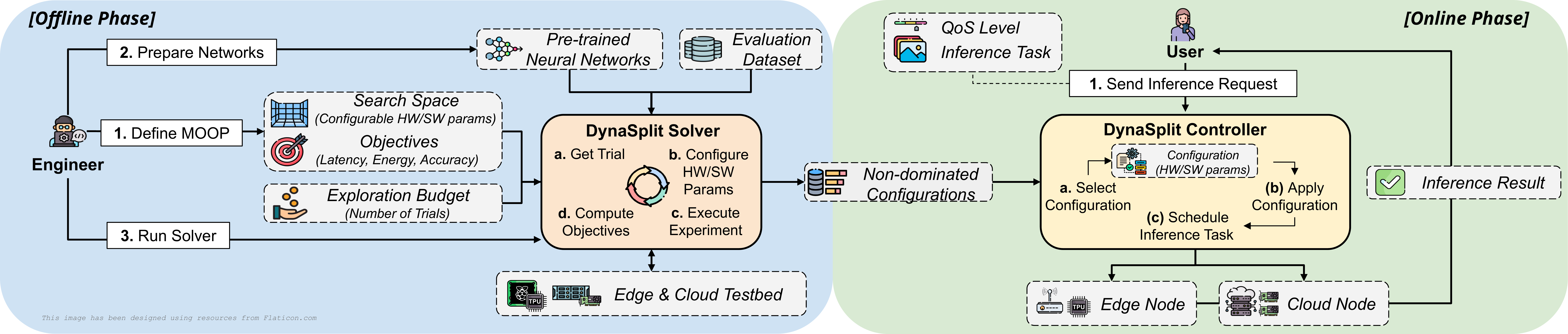}
    \caption{An overview of the \dynasplit framework.}
    \label{fig:overview}
\end{figure*}

\subsection{Optimization Problem}\label{sec:optimization-objective}

The primary objective of \dynasplit is to minimize latency and energy consumption while maximizing inference accuracy by leveraging a \textit{hardware-software co-design framework}.

This is formulated as a MOOP, defined as:
\begin{equation}
\label{eq:min-func}
    \text{minimize}_{x \in X} \left( T_{\text{inf}}(x), E_{\text{inf}}(x), -A(x) \right)
\end{equation}
Where $A(x)$ represents the accuracy of the inference for a given configuration $x$ that maps from the search space (i.e., configuration space) $X$ to an accuracy value.

Solving the MOOP results in a set of solutions known as the \emph{Pareto front}. This Pareto front consists of \emph{non-dominated configurations}, i.e., no other configuration in the search space can improve one objective without worsening at least one other objective~\cite{ngatchou2005pareto}. In other words, the Pareto front captures suitable trade-offs among target objectives, allowing  different configurations that balances the objectives.

\section{\dynasplit: System Overview \& Workflow}\label{sec:approach}

We propose \dynasplit, an energy-aware approach that combines both software parameters (e.g., NN split layer) and hardware parameters (e.g., accelerator usage, CPU frequency) in a \emph{co-design optimization framework} to efficiently serve inference requests at the edge.

\subsection{System Overview}

Figure~\ref{fig:overview} illustrates the \dynasplit framework, comprising two main components: the \solver and the \controller, which manage the \emph{Offline Phase} and \emph{Online Phase}, respectively.

The \solver solves the MOOP by efficiently exploring the hardware-software search space. It uses a meta-heuristic optimization algorithm to identify configurations for model split points and edge/cloud nodes, suitable for large search spaces with costly optimization objectives~\cite{DBLP:conf/kbse/TundoMIBBM23}.

The \controller manages incoming inference requests by selecting the most energy-efficient configuration from the \solver's solutions. It ensures the chosen configuration meets the required QoS level, respecting the inference latency threshold while minimizing energy consumption without significantly affecting accuracy.

\subsection{Offline Phase}
The \emph{Offline Phase} consists of three tasks. Firstly, we must define the MOOP by providing the optimization objectives and the relevant hardware and software parameters that comprise the search space. Secondly, it is essential to ensure that pre-trained NNs are prepared to accommodate all potential combinations of head and tail models. Finally, the engineer can run the \solver to explore the search space and determine the non-dominated configurations.

\subsubsection{Defining the MOOP}
\dynasplit optimizes - but is not limited to - three objectives as defined in \S~\ref{sec:optimization-objective}, that is, minimizing the energy consumption, minimizing the inference latency, and maximizing the inference accuracy. The search space $X$ contains hardware and software parameters, and it is defined as a set of configuration tuples. The search space $X$ we consider contains four parameters, that is, the CPU frequency of the edge node ($CPU_f$), the frequency of the edge hardware accelerator ($TPU_f$), the usage of a cloud GPU, and the NN split layer ($L_{\{network\}}$). Each parameter domain has a different cardinality (see details in Table \ref{tab:search-space}). Accordingly, $|X| = |CPU_f| \times |TPU_f| \times |GPU| \times |L_{\{network\}}|$ configuration tuples. For example, for the VGG16 network $|X| = 7 \times 3 \times 2 \times 23 = 966$.

However, while the initial search space $X$ includes all possible parameter combinations, some configurations are infeasible because of the conditional nature of our search space; that is, the possibility to assign a certain value $v$ to a parameter $p_1$ depends on the assigned value $w$ to a parameter $p_2$. In particular,
\begin{enumerate*}[label=(\roman*)]
    \item \textit{TPU usage with cloud-only inference ($k = 0$):} No TPU is used when all the computation is performed on the cloud node, as there is no edge processing involved. 
    \item \textit{GPU usage with edge-only inference ($k = L$):} No GPU is used when all the computation is performed on the edge node, as there is no cloud processing involved.
\end{enumerate*}

In addition, each NN might introduce further constraints due to technical limitations in its architecture and hardware dependencies, requiring us to define a search space for each network. For instance, among our two evaluated  pre-trained NNs, that is, VGG16~\cite{vgg} and Vision Transformer (ViT)~\cite{vit}, 
for the ViT network, the edge TPU is not used in any configuration due to specific memory limitations~\cite{DBLP:conf/wmcsa/WangZWY22}; therefore, all the configurations containing the $TPU_f$ parameter set to 250 or 500 are considered unfeasible.

\subsubsection{Preparing the Networks}\label{sec:preparing-networks}

Model partitioning varies depending on the specific neural network architecture, leading to different potential splitting points for each network. For example, VGG16 allows partitioning after each of its 22 layers, while ViT can be partitioned at various points within its architecture.

To execute on specialized hardware like edge TPUs, network layers must be prepared, that is, head portions require 8-bit integer quantization and  compilation for TPU execution. This post-training quantization process may slightly affect accuracy~\cite{DBLP:journals/air/ChoudharyMGS20,DBLP:journals/pieee/DengLHSX20}, which we address by including accuracy as an optimization objective. For larger networks like ViT~\cite{DBLP:conf/wmcsa/WangZWY22}, which are too large for edge  accelerator (TPU) quantization, head portions can be executed using standard 32-bit floating point operations on the edge node's CPU.

\subsubsection{Running the \dynasplit Solver}

To solve the MOOP, \dynasplit leverages NSGA-III~\cite{nsgapart1, nsgapart2}, a robust and fast multi-objective optimization algorithm that has been successfully applied to several application domains~\cite{DBLP:conf/gecco/Fieldsend17,DBLP:conf/gecco/MarkoPCD19,DBLP:conf/gecco/FitzgeraldDHM22}. It employs a set of predefined reference points to direct the search process to  maintain the diversity in the solutions, which is important when optimizing more than two objectives, as it prevents the algorithm from clustering candidate configurations in specific regions of the search space~\cite{nsgapart1, nsgapart2}.

For each candidate configuration to test (i.e., a trial), the \solver configures the edge-cloud testbed accordingly. This includes loading the head and tail networks and setting all the hardware parameters included in the search space to the specific configuration values. Then, it executes the inference task using samples retrieved from the evaluation data set. During the execution of the experiment, it collects and stores the objective values (i.e., inference latency, inference energy consumption, and accuracy), which are, in turn, used by the optimization algorithm to evaluate the quality of the solution.

The \solver records all the objective values evaluated throughout its operation. Upon completion, it extracts the non-dominated configuration set from the complete set of results. Our empirical evaluation shows that exploring 20\% of the search space is sufficient to identify suitable configurations when compared to a larger search of $\sim80\%$. This is particularly significant because evaluating how a single configuration satisfies the objectives requires gathering several empirical measurements. %

In fact, to obtain reliable performance metrics, \dynasplit averages results over 1,000 inferences for each configuration. This helps to capture the fluctuations in the testbed and  significant number of samples to be collected for metrics whose sampling rate is slower than the single inference time due to physical limitations (i.e., power meters). %

\subsection{Online Phase}
The \emph{Online Phase} begins with the \controller receiving a user's request containing the inference task (e.g., object detection in a set of images) and the requested QoS level expressed as the maximum acceptable inference latency in milliseconds. The QoS level can be obtained, for example, from Service Level Agreements (SLAs), application class (i.e., time-critical vs. non-time critical applications), or mobile network connection type (e.g., LTE vs. 4G vs. 5G). When it receives the request, the \controller performs three main tasks detailed hereafter: (i) selecting the most energy efficient - but yet suitable - configuration to meet the requested QoS; (ii) applying the selected configuration by tweaking both software and hardware settings of the edge and cloud nodes; and (iii) executing the inference by ultimately scheduling the requested task.

\subsubsection{Selecting Configuration}
The \controller uses non-dominated configurations identified during the \emph{Offline Phase} to select the most suitable configuration for an incoming request. %
 At the  start of the system, it sorts and keeps in memory the non-dominated configuration set by using the following sort criteria: (i) ascending energy consumption (the lower the better), and (ii) descending accuracy (the higher the better).

\begin{algorithm}[!ht]
\caption{Request Scheduling and Configuration}
\label{alg:runtime_scheduling}
\begin{algorithmic}[1]
\Require $qos$, the QoS level expressed as maximum inference latency (ms)
\Require $sortedConfigSet$, the sorted non-dominated configuration set
\State $config \gets sortedConfigSet[0]$ 
\For{$i \gets 0, \text{size}(sortedConfigSet)$}
    \If{$sortedConfigSet[i].\text{latency} \leq qos$}
        \State \textbf{return} $sortedConfigSet[i]$  
    \EndIf
    \If{$sortedConfigSet[i].\text{latency} < config.\text{latency}$}
        \State $config \gets sortedConfigSet[i]$ 
    \EndIf
\EndFor
\State \textbf{return} $config$  
\end{algorithmic}
\end{algorithm}

Algorithm~\ref{alg:runtime_scheduling} shows the pseudo-code for our configuration selection algorithm. Initially, it selects the first configuration present in the sorted set (line 1), that corresponds to the most energy efficient as result of the sorting operation. Afterward, it searches for the \emph{most energy efficient configuration that satisfies the required QoS level}, that is, a configuration whose inference time is less than or equal to the QoS level. If such a configuration is found, it is selected (lines 2 - 4). Otherwise, \dynasplit continues to search for the \emph{fastest available configuration}, even if it exceeds the QoS level (lines 6 - 10). This ensures that the system provides the best possible performance, trying to minimize QoS violations. %
The algorithm's runtime complexity is $O(n)$, where $n$ represents the size of the sorted set of non-dominated configurations. In the worst case, the algorithm must iterate through the entire set once for each request in order to identify a suitable configuration.

\subsubsection{Applying Configuration}
Applying the selected configuration requires adjusting settings on both the edge and cloud nodes. At the edge node, the \controller first adjusts both the CPU and the TPU frequencies to the selected configuration values. In particular, when $TPU_f = off$, the edge hardware accelerator is not used; the TPU is completely turned off to avoid energy waste. Moreover, the head network is loaded when not previously in use.

When $config.L_{network} \ne L$, and so the inference task will also use cloud computation, the \controller sends an initialization message to the cloud node indicating the tail network to load and also whether to use the GPU for cloud acceleration. %

\subsubsection{Executing Inference}
The inference task can be scheduled at this stage. Here, the head network processes the user's data, producing intermediate results. Then, intermediate results are streamed to the cloud node, where the tail network processes them and streams the final results back to the edge node. Finally, inference results are forwarded to the requesting user, terminating the request cycle.

\section{Implementation}\label{sec:implementation}
\dynasplit has been prototyped using Python. 

Our \solver leverages Optuna~\cite{optuna_2019}, an open-source hyperparameter optimization framework with MOOP capabilities. The optimization procedure employs the NSGAIIISampler~\cite{nsgaiii_optuna} implementing the NSGA-III algorithm with default parameter values, and the result database provided by Optuna. We also use the GridSampler~\cite{gridsearch_optuna} for the larger exploration of the search space conducted in our empirical evaluation.

The communication between the edge and cloud nodes is implemented using gRPC~\cite{grpc} with bidirectional streaming. The bidirectional streaming sends metadata only once at the beginning of the stream, rather than with each individual request. This method allows us to efficiently transmit data in a continuous stream, reducing memory overhead by clearing intermediate outputs progressively, a crucial aspect given the limited resources available on the edge node.

Our prototype supports two NNs pre-trained on ImageNet dataset~\cite{ILSVRC15}, namely VGG16~\cite{vgg} and Vision Transformer (ViT)~\cite{vit}. VGG16 is implemented via TensorFlow's Keras applications, while ViT uses an alternative Keras implementation~\cite{faustomorales2023}.

As described in \S~\ref{sec:preparing-networks}, NNs necessitate of preliminary steps (e.g., splitting in head and tail portions) to allow their execution. In particular, the head portions of VGG16 network are quantized to 8-bit integer and compiled specifically for the TPU using LiteRT~\cite{litert}, formerly Tensorflow Lite, to execute inference on the edge node using the TPU. Quantization is performed using 100 random images from the ImageNet validation dataset. Due to memory constraints~\cite{DBLP:conf/wmcsa/WangZWY22}, ViT does not run on the edge TPU and is instead processed with standard 32-bit floating point operations. In addition, the ViT network requires an additional step to approximate the iGELU activation function, as TensorFlow Lite does not support it~\cite{Reidy2023}. Inference on the cloud node is performed without any modifications to the tail portions in a TensorFlow GPU-ready Docker container~\cite{tensorflow_gpu_container}.

\section{Empirical Evaluation}\label{sec:empirical-evaluation}

We empirically evaluate \dynasplit by employing both simulations and experiments on a realistic edge-cloud testbed. We describe hereafter the experimental setup (\S~\ref{sec:experimental-setup}), the experimental plan (\S~\ref{sec:experimental-plan}), the empirical results (\S~\ref{sec:testbed-results}, \ref{sec:simulation-results}, \ref{sec:overhead}), and we conclude with a final discussion (\S~\ref{sec:discussion}).

\subsection{Experimental Setup}\label{sec:experimental-setup}

\begin{figure}[!ht]
    \centering
    \includegraphics[width=\linewidth]{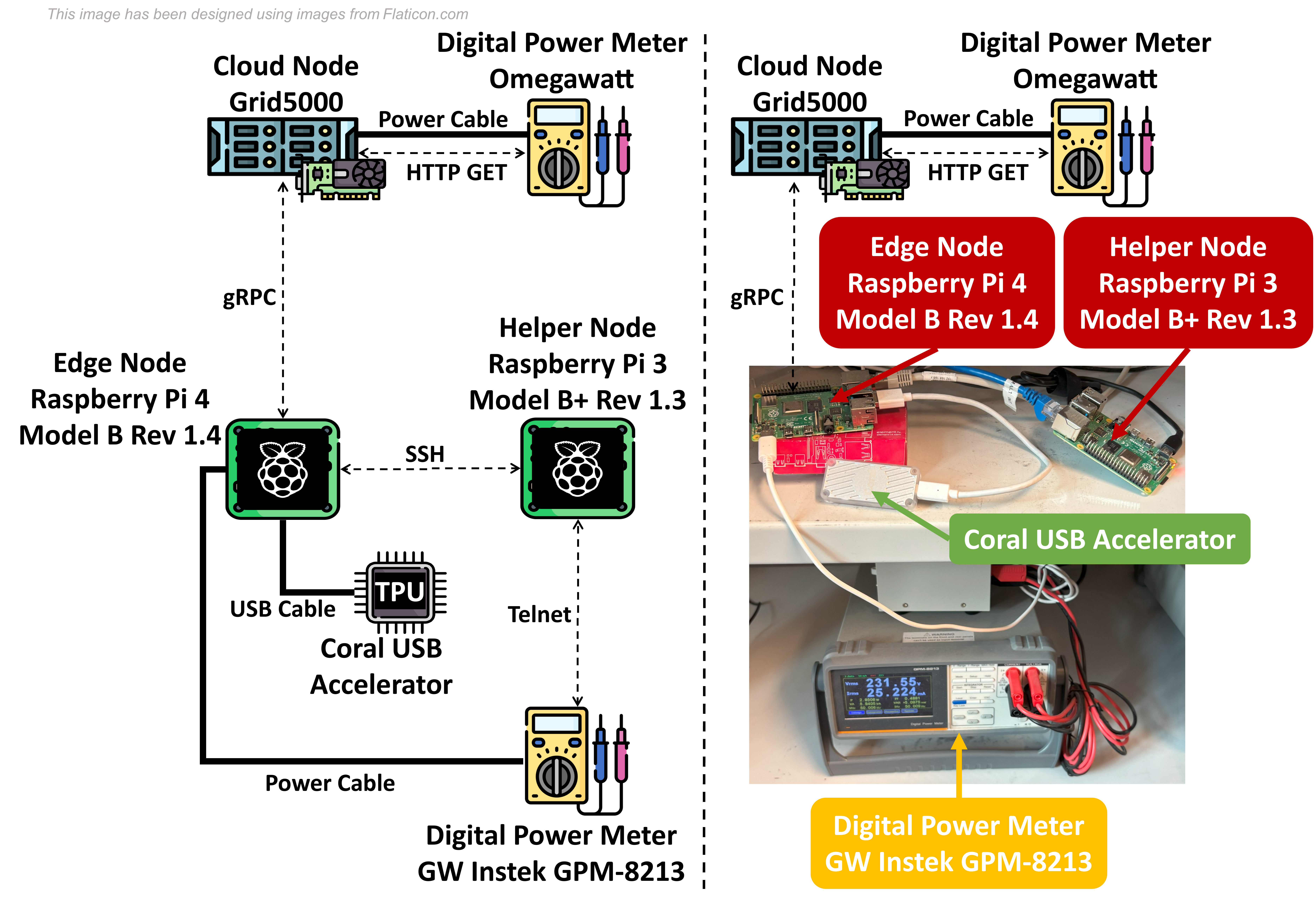}
    \caption{The testbed used to run the empirical evaluation.}
    \label{fig:testbed}
\end{figure}

Fig.~\ref{fig:testbed} presents the testbed used for our empirical evaluation, with a schematic overview on the left and the in-lab implementation on the right.

\textbf{Edge Node}. The edge node runs in our laboratory and consists of a Raspberry Pi 4 Model B Rev 1.4 (Raspberry Pi OS 64-bit, 8 GB of RAM, quad-core ARMv8). We disable unnecessary components such as WiFi, Bluetooth, and both the power and activity LEDs to reduce power consumption. Normally, the CPU frequency dynamically scales between 600 MHz and 1.5 GHz, with throttling engaged if the CPU temperature exceeds 80-85°C~\cite{rpi_throttling}. However, we set the CPU frequency governor to \texttt{userspace}, allowing us to avoid dynamic scaling and explicitly set the CPU frequency (from 0.6GHz to 1.8GHz with 0.2GHz increments). %
The edge node is equipped with a Google Coral USB Accelerator~\cite{coral_accelerator} (4 TOPS int8 TPU coprocessor). The TPU operates at 250 MHz using the \texttt{libedgetpu1-std} package or at 500 MHz with \texttt{libedgetpu1-max}. We disable the USB port when the TPU is not required, preventing it from drawing idle power.
We measure the active power of the edge node with a GW Instek GPM-8213 Digital Power Meter~\cite{gpm_power_meter}, which provides a sampling rate of 200 ms and a resolution of 1 mW. Energy consumption is then calculated using trapezoidal integration over the sampled power data.

\textbf{Cloud Node}. The cloud node is part of the Grid5000 cluster~\cite{grid5k}, and is located at a completely different site from the edge node. The node is equipped with dual Intel Xeon E5-2698 v4 CPUs, 512 GiB RAM, and 8 NVIDIA Tesla V100 GPUs, although we restrict using a single GPU in our experiments. The node is connected to a physical Omegawatt wattmeter, which provides power measurements at 20 ms intervals with a precision of 0.1 W. The ability to directly measure power consumption at the node level makes this node suitable for our edge-cloud testbed.

\textbf{Helper Node}. We use a third node in our experimental setup to run the \dynasplit components, that is, the \solver and the \controller. The node consists of Raspberry Pi 3 Model B+ Rev 1.3 (Raspberry Pi OS 64-bit, 1 GB of RAM, quad-core ARMv8).

\subsection{Experimental Plan}\label{sec:experimental-plan}
We conduct two experiments, namely \emph{Testbed Experiment} and \emph{Simulation Experiment}. Both the experiments use the same set of evaluation metrics described in \S~\ref{sec:evaluation-metrics} and compare \dynasplit with four baselines described in \S~\ref{sec:baselines}. 

In particular, the \emph{Testbed Experiment} assesses \dynasplit in a realistic deployment by employing our edge-cloud testbed and handling a workload of 50 user requests per network as described in \S~\ref{sec:workload}. In this experiment, we also compare our evaluation metric results when \dynasplit utilizes the set of non-dominated configurations obtained through the exploration of 20\% of the search space with our \solver with the set of non-dominated configurations obtained through a larger search of $\sim80\%$.

On the other hand, the \emph{Simulation Experiment} aims to assess \dynasplit when handling a higher number of user requests. In this case, we simulate up to 10,000 user requests by reusing the evaluation metrics obtained during the search space exploration and the \emph{Testbed Experiment}. We ensured each configuration used in the simulation was evaluated at least five times on the testbed and randomly sampled from the pool of observations for given configurations.

Finally, we also conduct an analysis to assess the run-time overhead of the \controller (\S~\ref{sec:overhead}).

\subsubsection{Workload Generation}\label{sec:workload}
We generate 50 requests for the \emph{Testbed Experiment} and 10,000 requests for the \emph{Simulation Experiments}. Each request represents a user who requires an object detection inference task with 1,000 images from the ImageNet validation dataset.

\begin{table}[!ht]
\centering
\caption{The latency upper and lower bounds for the VGG16 and ViT networks.}
\label{tab:latency_bounds}
\resizebox{\columnwidth}{!}{%
\begin{tabular}{l|rl|rl}
                   & \multicolumn{2}{c}{\textbf{Min. Latency}}                                                                                                                      & \multicolumn{2}{c}{\textbf{Max. Latency}}                                                                                                        \\
                   & \multicolumn{1}{c}{\textit{Value}} & \multicolumn{1}{c}{\textit{Configuration}}                                                                                      & \multicolumn{1}{c}{\textit{Value}} & \multicolumn{1}{c}{\textit{Configuration}}                                                                        \\ \hline
VGG16              & 90.6 ms                        & \begin{tabular}[c]{@{}l@{}}CPU Freq.: 1.2 GHz\\ TPU: No\\ GPU: Yes\\ Split Layer: 0\end{tabular}                 & 5,026.8 ms                      & \begin{tabular}[c]{@{}l@{}}CPU Freq.: 0.6 GHz\\ TPU: No\\ GPU: No\\ Split Layer: 20\end{tabular} \\ \hline
ViT & 118.8 ms                        & \begin{tabular}[c]{@{}l@{}}CPU Freq.: 1.4 GHz\\ TPU: No\\ GPU: Yes\\ Split Layer: 0\end{tabular} & 10,287.6 ms                     & \begin{tabular}[c]{@{}l@{}}CPU Freq.: 0.6 GHz\\ TPU: No\\ GPU: No\\ Split Layer: 18\end{tabular}
\end{tabular}%
}
\end{table}

\begin{figure}[!ht]
    \centering
    \includegraphics[width=0.6\linewidth]{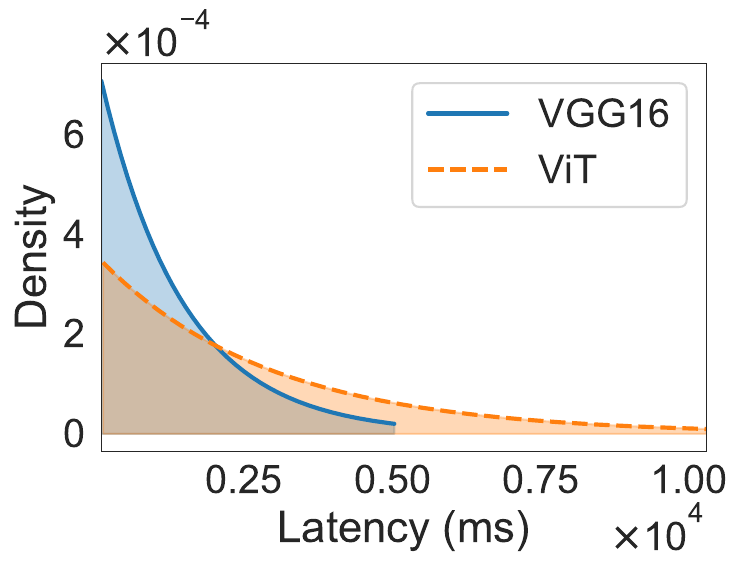}
    \caption{Inference time request distributions for VGG16 and ViT networks.}
    \label{fig:inference_requests}
\end{figure}

To assign a QoS level to each request, we use the Weibull distribution with the shape parameter set to 1 (i.e., it reduces to an exponential distribution), a well-known distribution that models real-world latency distribution~\cite{DBLP:journals/jnca/ArfeenPMW19}. For each neural network, we generate samples from the distribution, and we scale its samples so that the smallest value corresponds to the minimum observed latency, while the largest matches the maximum observed latency for the given network (see Table~\ref{tab:latency_bounds}). Figure~\ref{fig:inference_requests} illustrates the distribution of the generated latency values for the two networks.

\subsubsection{Evaluation Metrics}\label{sec:evaluation-metrics}
We evaluate \dynasplit using four metrics: latency, QoS violations, energy consumption, and inference accuracy.

\textbf{Latency}. We measure latency in milliseconds  using Python's \texttt{time.perf\_counter\_ns}. We compute total latency as the sum of:
\begin{enumerate*}[label=(\roman*)]
    \item \emph{Edge Latency} includes image scaling, batch creation, NN inference, and final output decoding;
    \item \emph{Cloud Latency} comprises deserialization of intermediate outputs, tail network invocation, and final output decoding;
    \item \emph{Network Latency} is the remaining time after accounting for edge and cloud latencies (i.e., data transfer time).
\end{enumerate*}

\textbf{QoS Violations}.
We determine QoS violations by counting requests where latency exceeds the specified QoS threshold and calculate the extent of the violation (in milliseconds).

\textbf{Energy Consumption}.
We track energy consumption in Joules separately for edge and cloud. To ensure stable measurements, 1,000 inferences per user request are batched. The edge performs 1,000 head inferences, sending results to the cloud for 1,000 tail inferences. This method allows for reliable power measurement by extending the inference period and overcoming the limitations of the power meter sampling rate (200 ms for the edge node and 20 ms for the cloud node).

\textbf{Accuracy}. %
It is defined as the ratio of correctly classified images to the total number of requests.

Please note, %
each user request consists of 1,000 inferences; the evaluation metric values for each request are calculated by averaging the results over these 1,000 inferences.

\subsubsection{Baseline Methods}\label{sec:baselines}
We examine \dynasplit's dynamic configuration selection by comparing it with the following four baselines, defined as follows. 
\begin{enumerate*}[label=(\roman*)]
    \item \textbf{Cloud-Only (cloud)}, that is, all inferences are processed on the cloud node using the GPU, with the edge CPU frequency set to the maximum value;
    \item \textbf{Edge-Only (edge)}, that is, all inferences are processed on the edge node, utilizing the TPU at its maximum frequency (or turning it off when not utilized as in the case of ViT) and setting the edge CPU frequency to its highest level;
    \item \textbf{Fastest (latency)}, that is, the fastest configuration  selected from the non-dominated configurations obtained during \emph{Offline Phase};
    \item \textbf{Energy-Saving (energy)}, that is, the most energy-efficient configuration selected from the non-dominated configurations obtained during \emph{Offline Phase}.
\end{enumerate*}

\subsection{Testbed Experiment Results}\label{sec:testbed-results}

\begin{figure}[!ht]
  \centering
  \includegraphics[width=0.5\columnwidth]{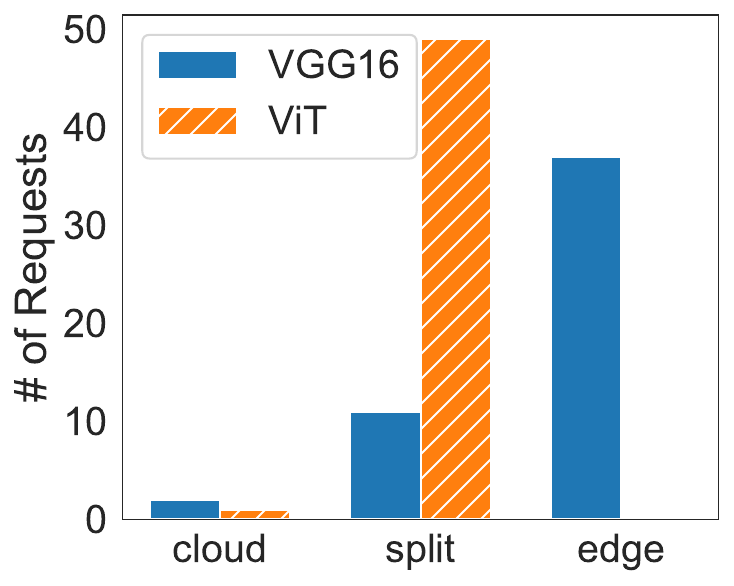}
  \caption{Scheduling decisions taken by \dynasplit.}
  \label{fig:scheduling_decision}
\end{figure}

Figure~\ref{fig:scheduling_decision} shows the scheduling decisions made by the \controller for each NN. For VGG16, 37 requests are scheduled for edge computation, only two for cloud, and 11 for split execution. For ViT, only one request is scheduled for cloud, with all remaining requests (49) using split computation. No edge computation is scheduled for ViT because the \solver did not identify any edge-only configuration during the \emph{Offline Phase}. VGG16's suitability for edge acceleration also explains its frequent edge computation use compared to ViT.

\subsubsection{Latency \& QoS Violations}

\begin{figure}[!ht]
  \centering
  \subfloat[VGG16]{\includegraphics[width=0.5\columnwidth]{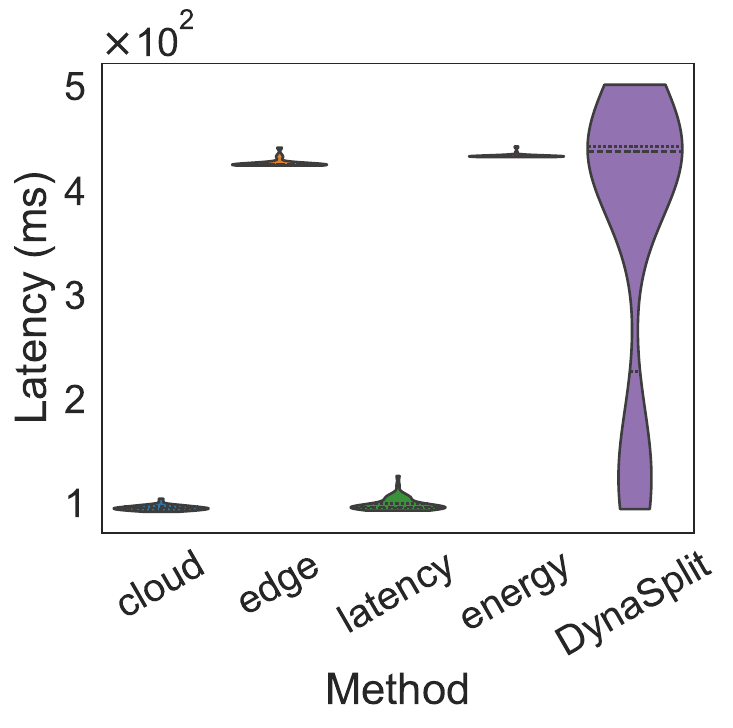}\label{fig:latency_vgg16}}
  \subfloat[ViT]{\includegraphics[width=0.5\columnwidth]{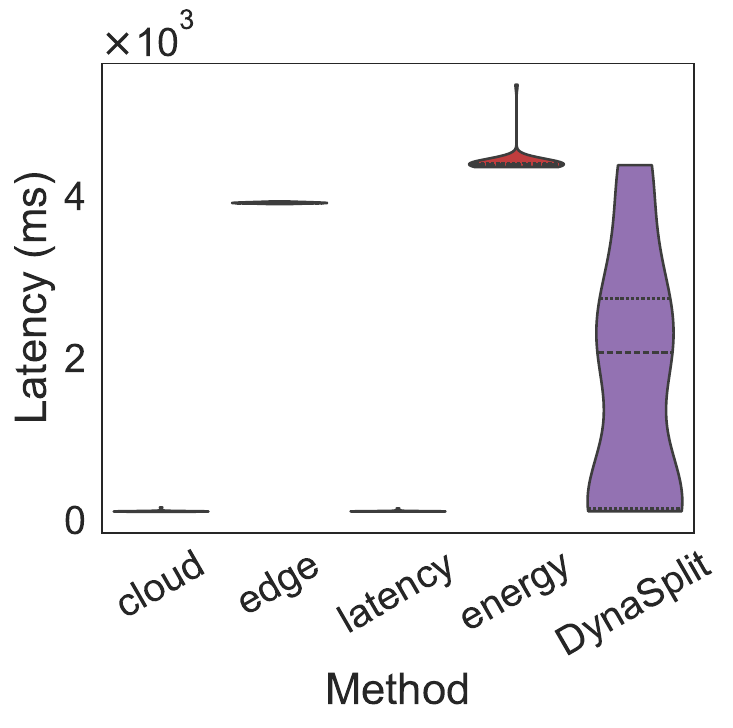}\label{fig:latency_vit}}
  \caption{Latency distribution for VGG16 and ViT networks.}
  \label{fig:latency}
\end{figure}

Figure~\ref{fig:latency} compares the latency distributions for 50 requests across four static baselines (cloud, edge, latency, energy) and \dynasplit. In all violin plots in our study horizontal lines represent quartiles, while the shape of the violin indicates the density. For VGG16, cloud and latency baselines reach a median of 96 ms and 97 ms respectively, while edge and energy baselines reach 425 ms and 434 ms, respectively. For ViT, cloud and latency report both a median of 117 ms, while edge and energy latency is 3,926 ms and 4,400 ms, respectively. \dynasplit adapts latency between edge and cloud, with a median close to edge for VGG16 and around 2,000 ms for ViT. Notably, for ViT, \dynasplit enables a finer-grained variety of latencies between edge and cloud latencies, compared to VGG16.

\begin{figure}[!ht]
  \centering
  \subfloat[VGG16]{\includegraphics[width=0.5\columnwidth]{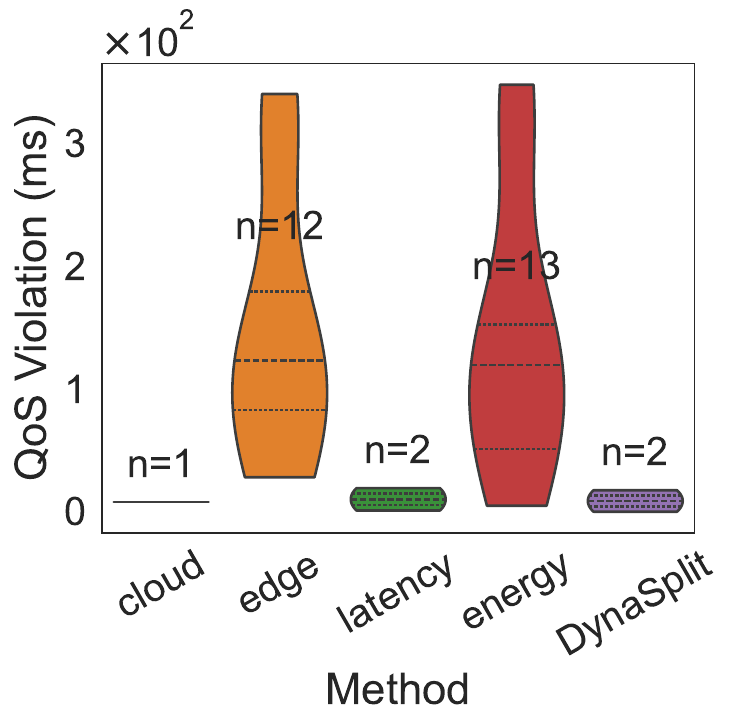}\label{fig:violation_vgg16}}
  \subfloat[ViT]{\includegraphics[width=0.5\columnwidth]{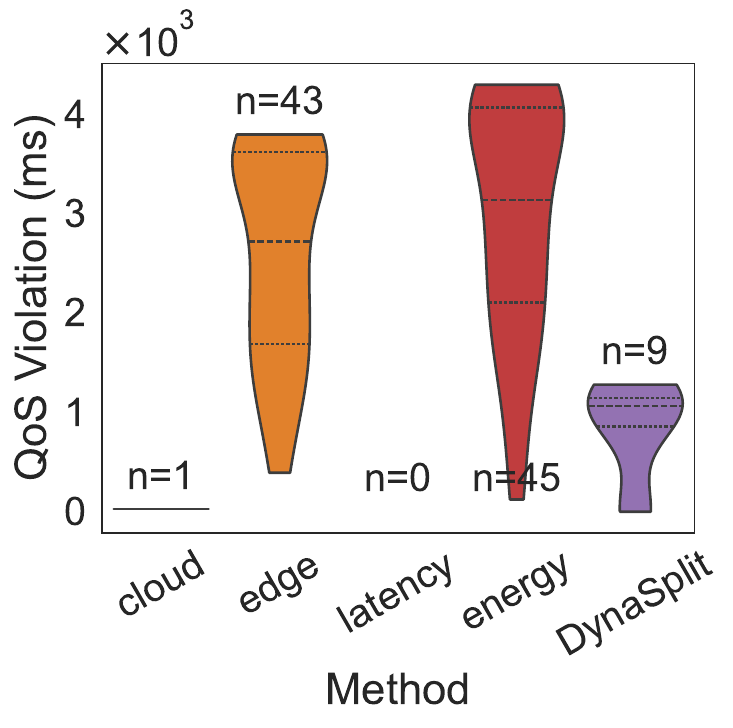}\label{fig:vialation_vit}}
  \caption{Distribution of QoS violations for VGG16 and ViT networks. The value $n$ indicates the number of violations.}
  \label{fig:violation}
\end{figure}

Figure~\ref{fig:violation} illustrates the distributions of QoS violations. Each violin plot shows how much the requests that missed their QoS deadlines exceeded the time limit. Both networks follow a similar pattern: the cloud and latency baselines violated the QoS level 2 times with minimal exceedance ($<20$ ms for VGG16, $<27$ ms for ViT). For edge and energy baselines, around 25\% of requests in VGG16 (up to 90\% in ViT) resulted in violations, with a median exceedance of 120 ms for VGG16 and roughly 3,000 ms for ViT. \dynasplit violates 4\% of the requests in VGG16 and 18\% in ViT, with a median exceedance of about 10 ms for VGG16 and 1000 ms for ViT.

\subsubsection{Energy Consumption}

\begin{figure}[!ht]
  \centering
  \subfloat[VGG16]{\includegraphics[width=0.5\columnwidth]{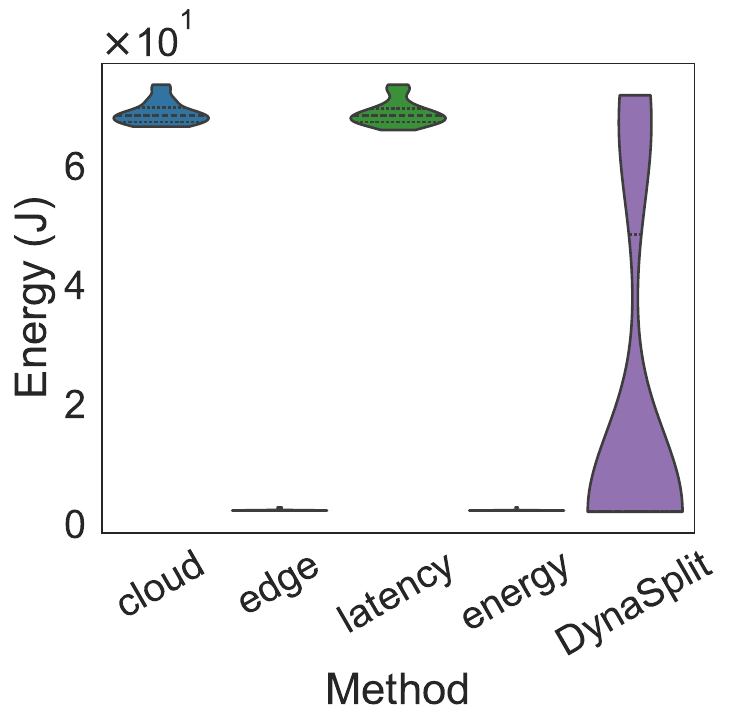}\label{fig:energy_vgg16}}
  \subfloat[ViT]{\includegraphics[width=0.5\columnwidth]{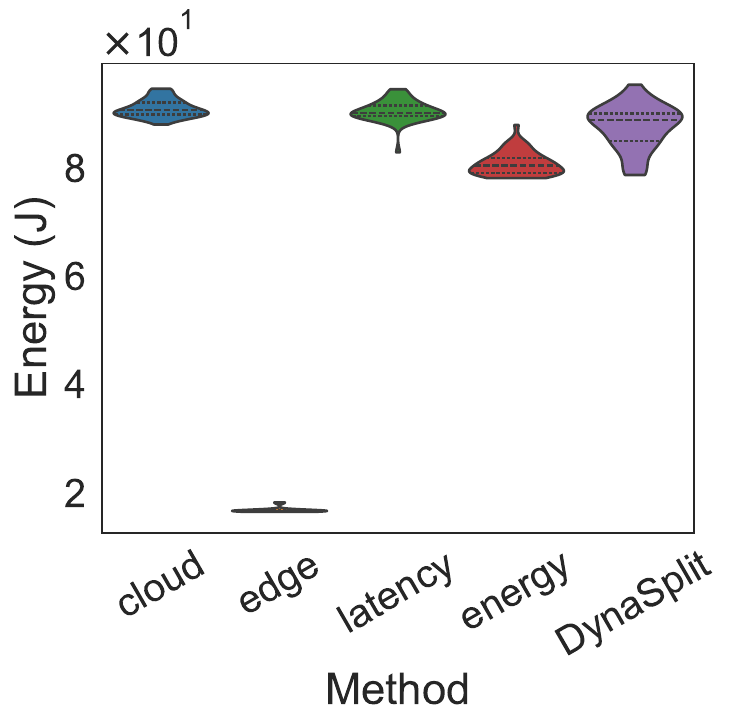}\label{fig:energy_vit}}
  \caption{Distribution of energy consumption for VGG16 and ViT networks.}
  \label{fig:energy}
\end{figure}

Figure~\ref{fig:energy} shows the energy consumption results. The cloud and latency baselines consistently consume more energy compared to other baselines and \dynasplit, reaching a median of $68$ J for VGG16 and $>90$ J for ViT. On the other hand, the edge and energy baselines for VGG16 consume in median $<3$ J. For ViT, the edge baseline consumes in median 16 J, while the energy baseline exhibits a higher median (80 J) because in this case the energy baseline (i.e., the most energy efficient in the non-dominated configurations) is a split configuration. \dynasplit dynamically adapts, with VGG16 showing a median energy comparable to the edge and energy baselines ($<3$ J), with the maximum reaching up to 72 J. For ViT, \dynasplit scores closer to cloud, energy, and latency baselines with a median of 89 J.

\subsubsection{Accuracy}

We observed negligible differences in accuracy ($< 1\%$). Cloud configurations demonstrate minimal advantages, most likely because they use the full model in FP32 without quantization. Nevertheless, our approach performs well both in terms of cloud accuracy and edge accuracy, demonstrating no negative impact when compared to baseline methods.

\subsubsection{\dynasplit Search vs. $\sim80\%$ Search}

We compare our \dynasplit approach, which explores 20\% of the search space (184 trials), with the results of exploring 81.5\% (80\% for the sake of simplicity) of the space (747 trials) for the VGG16 network. Empirical results show that the \controller scheduled an identical number of requests for cloud-only computation. For split computation and edge-only computation, we observed negligible differences, with only a single data point variation.

\begin{figure}[!ht]
  \centering
  \subfloat[Latency]{\includegraphics[width=0.333\columnwidth]{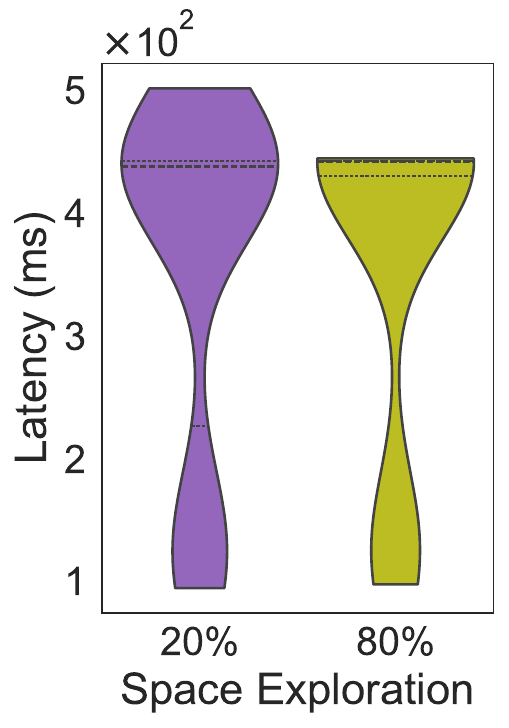}\label{fig:exhaustive_latency}}
  \subfloat[QoS Violation]{\includegraphics[width=0.333\columnwidth]{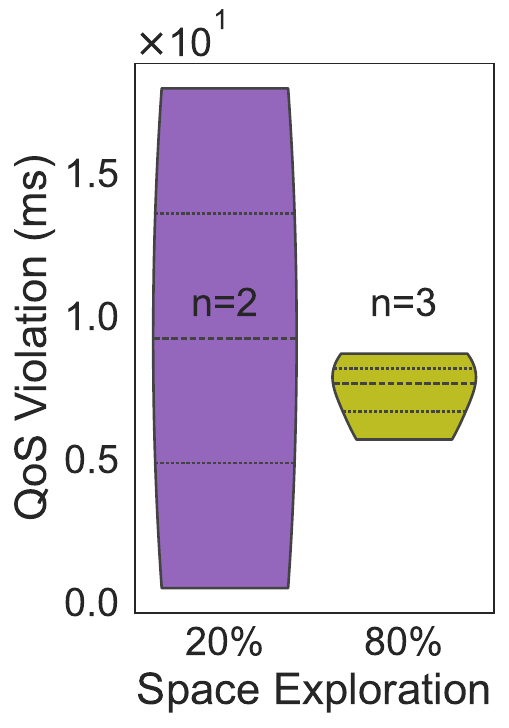}\label{fig:exhaustive_violation}}
  \subfloat[Energy]{\includegraphics[width=0.333\columnwidth]{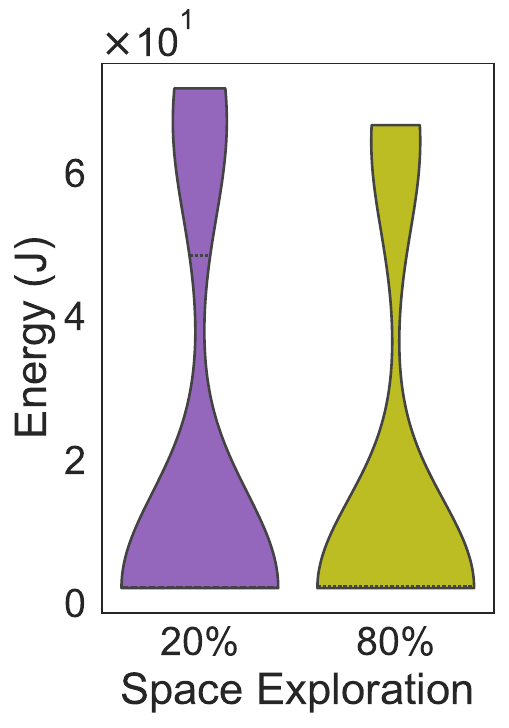}\label{fig:exhaustive_energy}}
  \caption{VGG16 latency, QoS violations, and energy consumption comparison between 20\% and 80\% exploration. %
  }
  \label{fig:exhaustive}
\end{figure}

We further analyzed how the exploration impacts the evaluation metrics. Figure~\ref{fig:exhaustive} shows that the two approaches produce very similar results, with no significant differences in latency, QoS violations, and energy consumption. In terms of latency (Figure~\ref{fig:exhaustive_latency}), both approaches handle the majority of requests within the expected timeframes. Although the 20\% exploration includes some latency observations near 500 ms, which the 80\% exploration does not, these occurrences are rare and do not significantly affect deadline fulfillment. Moreover, Figure~\ref{fig:exhaustive_violation} shows that both strategies maintain minimal QoS violations (less than 3), with violations missing a maximum of only 20 ms. Negligible differences are observed for energy consumption (Figure~\ref{fig:exhaustive_energy}).

To conclude, exploring only 20\% of the search space is sufficient, offering nearly identical performance to the 80\% exploration without noticeable shortcomings.

\subsection{Simulation Experiment Results}\label{sec:simulation-results}

\begin{figure}[!ht]
  \centering
  \includegraphics[width=0.5\columnwidth]{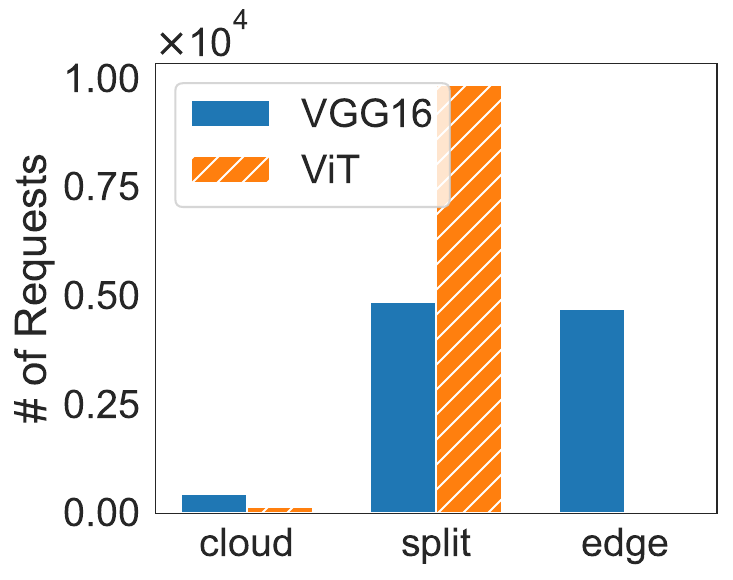}
  \caption{\dynasplit scheduling decisions during the \emph{Simulation Experiment} for the VGG16 and ViT networks.}
  \label{fig:scheduling_decision_scale}
\end{figure}

Figure~\ref{fig:scheduling_decision_scale} illustrates how \dynasplit schedules incoming requests across cloud, split, and edge computation for the VGG16 and ViT networks when simulating up to 10,000 requests. Both networks schedule only a small fraction of requests for cloud computation (4\% for VGG16 and 1\% for ViT), consistent with the \emph{Testbed Experiment}. When using ViT, no requests are scheduled as edge-only computation due to the absence of edge-only configurations in the non-dominated configuration set. However, VGG16 schedules more split decisions, likely due to the finer-grained variation of QoS levels, allowing the \controller to utilize a wider range of configurations. As a result, split and edge configurations share about half of the scheduling choices, 4857 and 4695, respectively.

\subsubsection{Latency \& QoS Violations}

\begin{figure}[!ht]
  \centering
  \subfloat[VGG16]{\includegraphics[width=0.5\columnwidth]{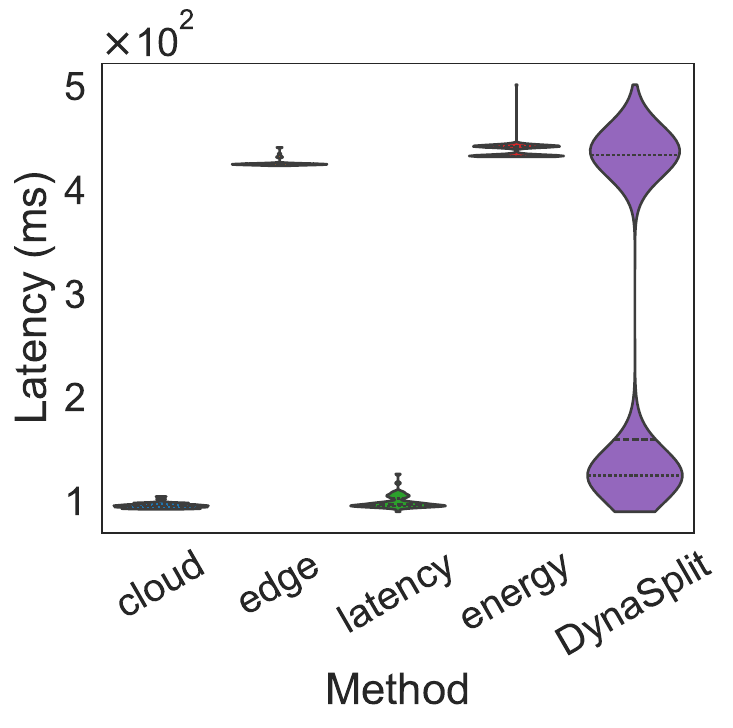}\label{fig:latency_vgg16_scale}}
  \subfloat[ViT]{\includegraphics[width=0.5\columnwidth]{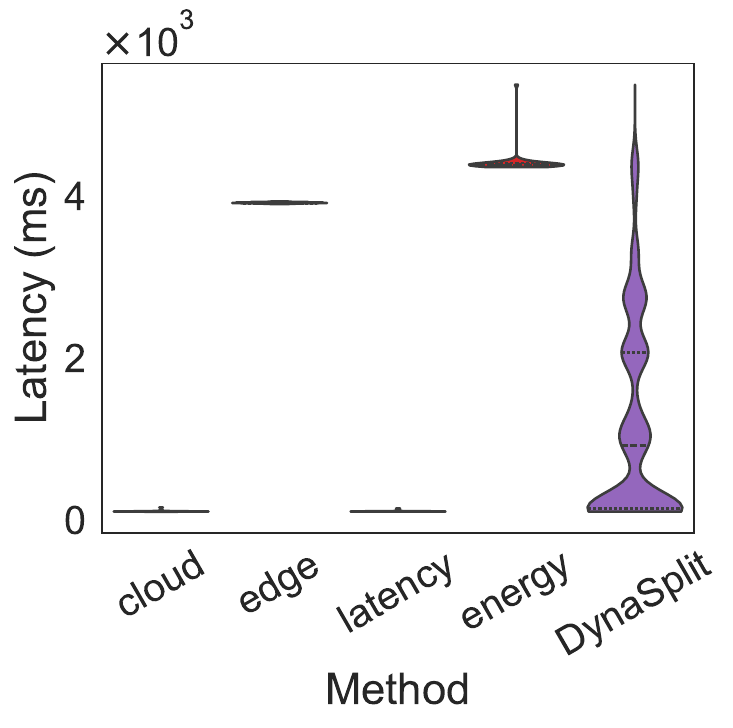}\label{fig:latency_vit_scale}}
  \caption{Latency distribution for the VGG16 and ViT networks in the \emph{Simulation Experiment}.}
  \label{fig:latency_scale}
\end{figure}

Figure~\ref{fig:latency_scale} shows the latency distribution for VGG16 (Figure~\ref{fig:latency_vgg16_scale}) and ViT (Figure~\ref{fig:latency_vit_scale}). Consistently with the \emph{Testbed Experiment}, cloud and latency baselines exhibit lower latencies (median of $96-97$ ms for VGG16, $117-118$ ms for ViT) when compared to edge and energy baselines (median of ($425-442$ ms for VGG16, $3926-4400$ ms for ViT). \dynasplit shows a partitioned distribution for VGG16 between cloud and edge latencies, with a lower median of 160 ms due to more requests being scheduled for split computation. For ViT, \dynasplit distributes in the range of edge and cloud baselines, with a high density at cloud latencies and a median of 933 ms.

\begin{figure}[!ht]
  \centering
  \subfloat[VGG16]{\includegraphics[width=0.5\columnwidth]{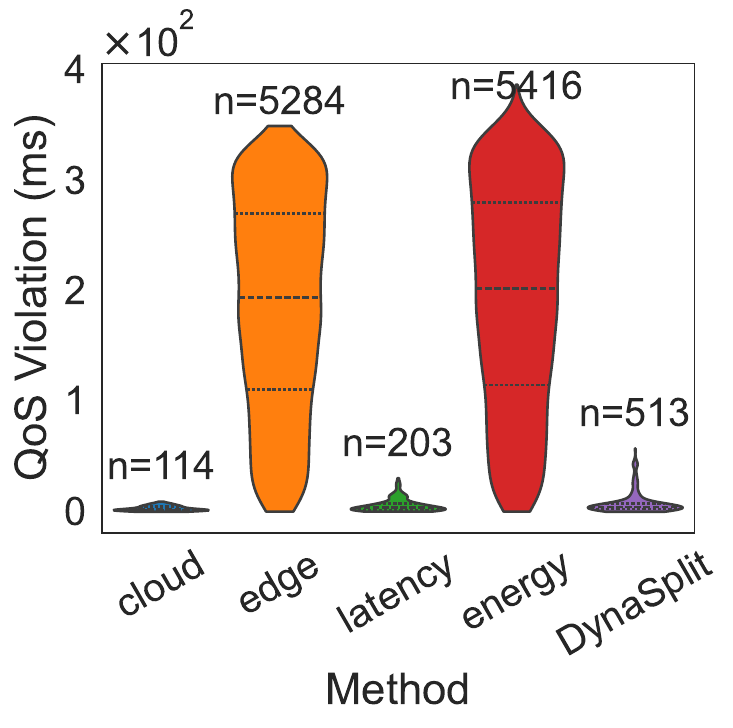}\label{fig:violation_vgg16_scale}}
  \subfloat[ViT]{\includegraphics[width=0.5\columnwidth]{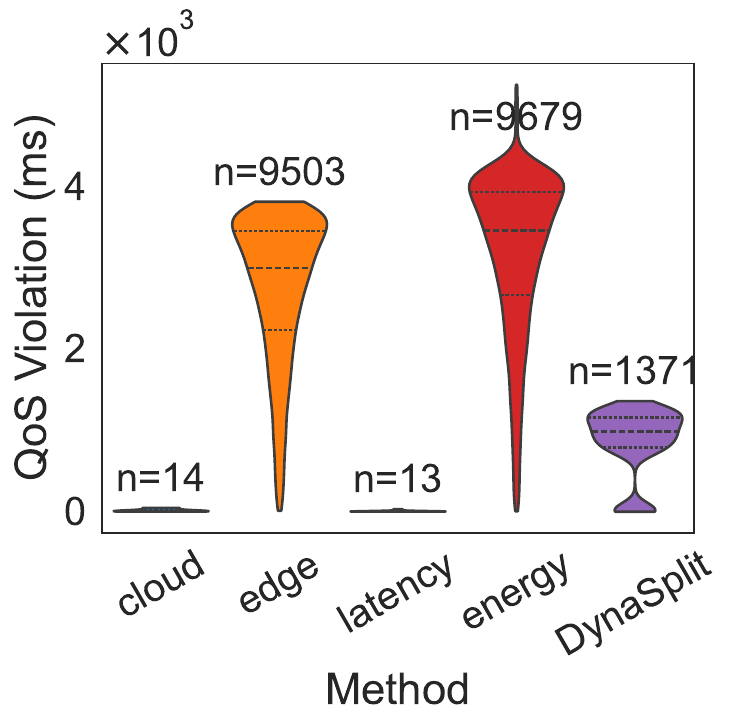}\label{fig:violation_vit_scale}}
  \caption{Distribution of QoS violations for VGG16 and ViT networks during the \emph{Simulation Experiment}.}
\label{fig:violation_scale}
\end{figure}

With respect to QoS violations shown in Figure~\ref{fig:violation_scale},  the cloud and latency baselines exhibit a small fraction of violations (at most 2\% for VGG16, and only 0.1\% for ViT), with median exceedances up to 3 ms for VGG16 and 10 ms for ViT. In contrast, edge and energy baselines show up to 54\% violations for VGG16 and up to 96\% for ViT, with median exceedances up to 202 ms for VGG16 and 3467 ms ViT. \dynasplit shows around 5\% violations for VGG16 and 14\% for ViT, with median exceedances of 4 ms for VGG16 and 986 ms for ViT. These patterns are consistent with the results observed in the \emph{Testbed Experiment}.

\subsubsection{Energy Consumption}

\begin{figure}[!ht]
  \centering
  \subfloat[VGG16]{\includegraphics[width=0.5\columnwidth]{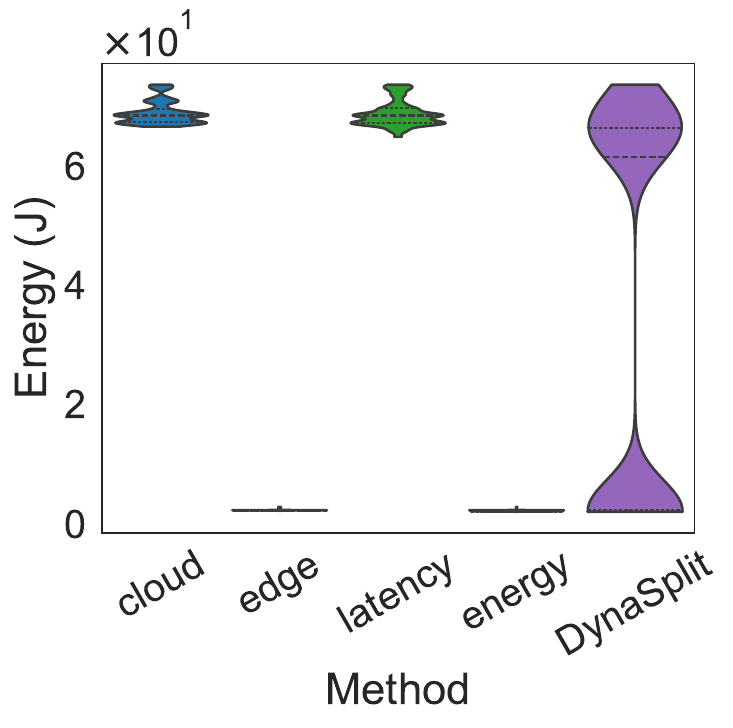}\label{fig:energy_vgg16_scale}}
  \subfloat[ViT]{\includegraphics[width=0.5\columnwidth]{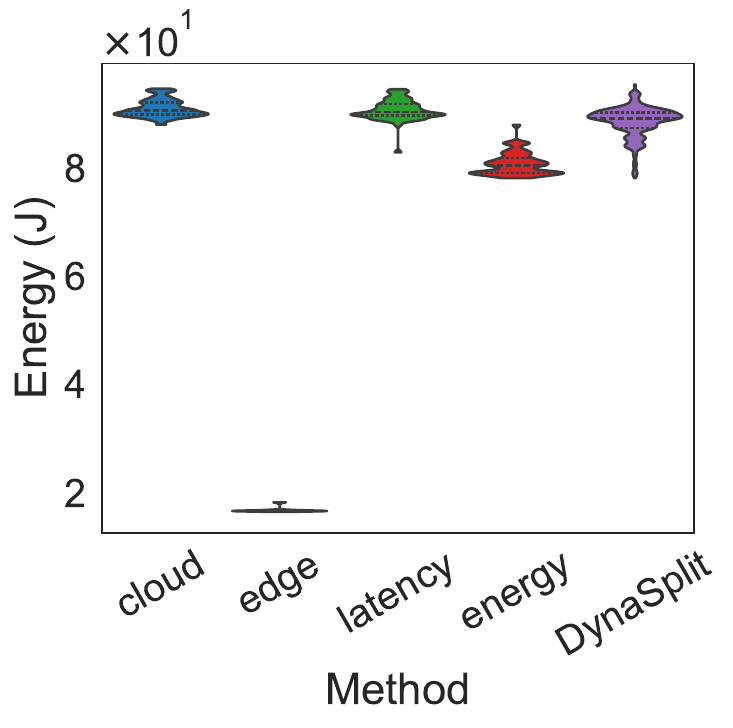}\label{fig:energy_vit_scale}}
  \caption{Energy consumption distribution for the VGG16 and ViT networks during the \emph{Simulation Experiment}.}
  \label{fig:energy_scale}
\end{figure}

As shown in Figure~\ref{fig:energy_scale}, the cloud and latency baselines consistently have high energy consumption (median of 69 J for VGG16, 91 J for ViT), while edge and energy baselines show lower energy  for VGG16 (median of 2 J). For ViT, due to the lack of an edge configuration in the offline phase, the energy baseline has a higher median of 81 J compared to the edge baseline (17 J). The \dynasplit approach for VGG16 demonstrates a distribution with the highest density at cloud and edge-like energy levels, with a median of 62 J, slightly higher due to more split decisions being scheduled. For ViT, \dynasplit shows cloud-like energy consumption with a median of 89 J, consistently showing a tendency to use less energy, aligning with the results from the \emph{Testbed Experiment}.

\subsubsection{Accuracy}

As for the \emph{Testbed Experiment}, also in the \emph{Simulation Experiment}, we observed negligible accuracy differences ($< 1\%$). \dynasplit performs at least as well as edge computation, maintaining comparable accuracy to the baselines without any significant impact.

\subsection{Overhead Analysis}\label{sec:overhead}
We analyzed the run-time overhead of the \controller by measuring the latency introduced by its operations and the memory usage. We collect such metrics while running the \emph{Testbed Experiment}.

On startup, the \controller loads and sorts the non-dominated configuration set only once. We observe that the median of loading and sorting time for this operation is 4.2 s, and uses up to 20 MB of memory. In comparison, loading the non-dominated configuration set that was obtained by searching $\sim80\%$ of the VGG16 search space took 29 seconds and resulted in a memory usage of 70 MB.

\begin{figure}[!ht]
  \centering
  \subfloat[Selecting Configuration]{\includegraphics[width=0.5\columnwidth]{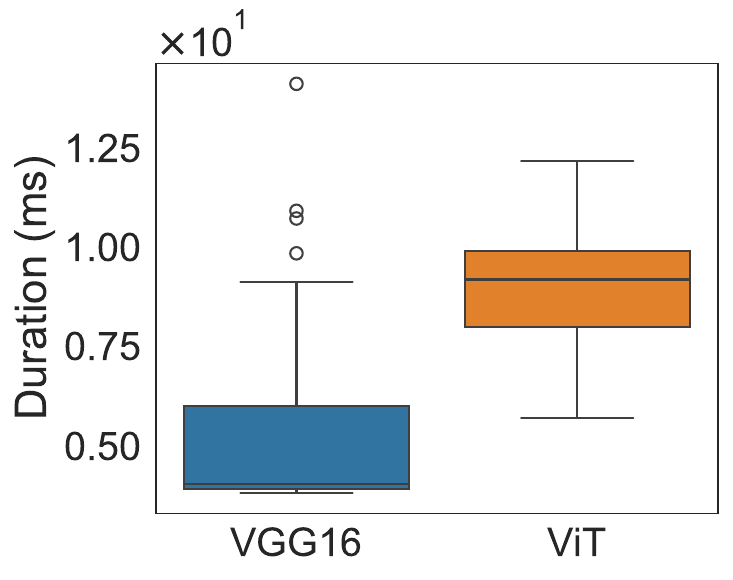}\label{fig:overhead_config_retrieve}}
  \subfloat[Applying Configuration]{\includegraphics[width=0.5\columnwidth]{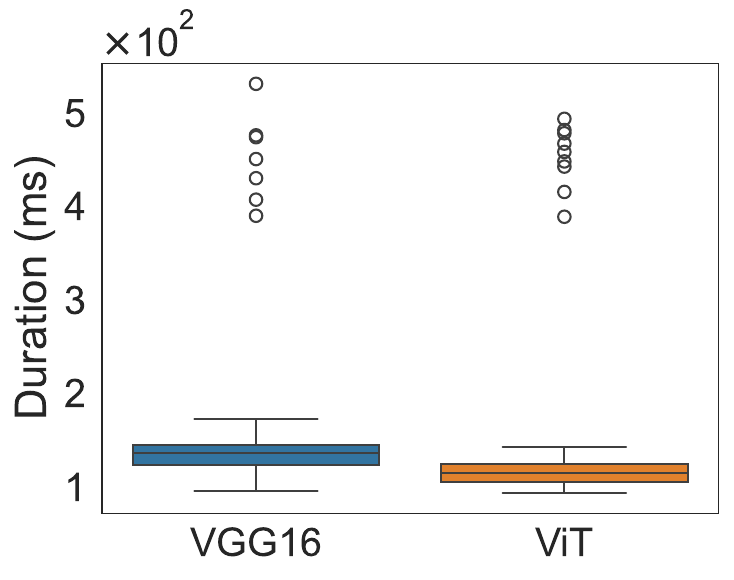}\label{fig:overhead_hardware_setup}}
  \caption{The overhead of the \controller for selecting the configuration and for applying the configuration for both VGG16 and ViT networks.}
  \label{fig:overhead}
\end{figure}

After the startup phase, the \controller only needs to select the most suitable configuration for the incoming request. As shown in Figure~\ref{fig:overhead_config_retrieve}, this takes up to 12 ms, with the median  taking less than 5 ms and 10ms for VGG16 and ViT, respectively. This duration is mainly influenced by the size of the non-dominated configuration set (in our case  15 and 12  for the two networks).

The most time-consuming operation is applying the selected configuration, as shown in Figure~\ref{fig:overhead_hardware_setup}. In most cases, this takes less than 200 ms, but we observed some outliers reaching up to 500 ms. However, the median for both the networks is below 150 ms.

We assess the overhead for the inference time by comparing the median overhead time to the median edge latencies. For VGG16, with a median edge latency of 426 ms, selecting the configuration increases the latency by 0.96\%, while applying the configuration increases the latency by 32.14\%. For ViT, with a median edge latency of 3922 ms, selecting the configuration increases the latency by 0.23\%, while applying the configuration increases the latency by 2.95\%.

These results show that while selecting the configuration introduces a minimal overhead, applying the configuration represents a significant part of the total overhead, though still relatively low compared to the overall inference time.

\noindent\textbf{Summary}. The empirical evaluation substantiates the efficacy of the proposed approach to scheduling inference requests in edge-cloud environments. By employing split computing techniques and optimizing relevant hardware parameters, we are able to effectively manage latency and energy consumption while maintaining accuracy, guaranteeing QoS requirements, and introducing minimal run-time overhead. Results show that \dynasplit can reduce energy consumption up to 72\% when compared to cloud-only computation, while meeting on average $\sim90\%$ of the request latency thresholds.

\subsection{Discussion}\label{sec:discussion}
We discuss here the potential shortcomings of our approach, highlighting conditions that may affect our findings from  real-world deployments. These considerations contextualize our results and suggest areas for future improvements.
\textbf{Offline Phase and Configuration Space Changes}.
Our approach uses offline optimization to find near-optimal hardware and software configurations. However, significant changes in model or hardware require re-optimization, as existing solutions may become obsolete. This dependency on offline optimization limits  adaptability, especially in dynamic environments where frequent updates to either models or hardware deployments could lead to additional costs and delays.

\textbf{Deployment Strategy}.
Our experiments use an always-on cloud server with pre-loaded models, ensuring minimal latency but potentially not reflecting real-world scenarios. Practical deployments often use on-demand services (i.e., serverless functions), which may incur cold-start latencies or additional bootup costs. While our setup prioritizes experimental consistency, future studies could explore more variable-resource environments to better represent typical deployment conditions.

\textbf{Overhead of Configuration Changes and Scheduling}.
Modifying hardware configurations (i.e., adjusting the CPU frequency or disabling the TPU) introduces processing overhead, but the precise time figures depend on the specific implementation (e.g., if DVFS can be applied straightforwardly or if it requires complex operations such as changing the OS libraries). Additionally, scheduling has a computational cost due to the fact that we have to search for suitable configurations for each request from a non-dominated configuration list. Although relatively small, this overhead impacts system efficiency and responsiveness, particularly in environments with strict latency requirements or high request volumes. One potential solution for this could be clustering user requests based on request type, QoS, and user profiles. This approach would reduce frequent configuration changes and decision overhead. Alternatively, using pre-determined configurations for specific request types could limit runtime adjustments, enhancing efficiency in real-world scenarios where rapid reconfiguration is challenging.

\section{Related Work}\label{sec:related-work}

Split computing approaches can be classified into two categories: \emph{with-mods} (modifying model architecture) and \emph{without-mods} (splitting at the desired layer). The latter approach profits from ability to divide NN models at intermediate layers whose output tensor size is smaller than the input~\cite{MatsubaraLR23}.

Among studies belonging to the \emph{with-mods} category~\cite{DBLP:conf/icassp/ChoiCB20,Li2020,DBLP:conf/mobicom/MatsubaraBCL019,DBLP:journals/access/MatsubaraCBLS20,DBLP:conf/islped/EshratifarEP19,DBLP:conf/sensys/Yao0LWLSA20,DBLP:conf/wowmom/MatsubaraCSLR22}, only a small number explicitly consider energy consumption as an evaluation metric or as a primary objective to be traded off with other application objectives~\cite{DBLP:conf/wowmom/MatsubaraCSLR22,DBLP:conf/sensys/Yao0LWLSA20}. Specifically, Matsubara\etal\cite{DBLP:conf/wowmom/MatsubaraCSLR22} propose a framework that integrates bottleneck injection with multi-stage training strategies,  to achieve enhanced accuracy even with more aggressive compression rates. However, all of the \emph{with-mods} approaches require greater investment in model engineering and retraining when compared to our approach.

As regards \emph{without-mods} approaches, most studies~\cite{Li2018,EshratifarAP21,PagliariCMP21,KangHGRMMT17,Luger2023,DBLP:conf/kdd/Banitalebi-Dehkordi21,DBLP:conf/www/ZhangC021,DBLP:journals/imwut/ZhangLLGWWDW20} evaluate various types of cost (e.g., computational load, energy consumption, communication cost) to partition models at each of their splitting points, while only a few choose splitting points based on heuristics~\cite{Choi2018,Cohen2020}.

Most of these works do not address application-specific performance metrics, such as accuracy, mainly because they don't modify input or intermediate representations~\cite{MatsubaraLR23}. In contrast, our approach considers accuracy, despite its minimal variance, as we employ quantization techniques for edge inference, which can slightly reduce model accuracy~\cite{DBLP:journals/pieee/DengLHSX20,DBLP:journals/air/ChoudharyMGS20}.

Some studies specifically focus on energy consumption as a primary objective, aligning with our approach. Kang\etal\cite{KangHGRMMT17} pioneered split computing with the Neurosurgeon framework, which finds optimal partitioning to minimize energy or latency using performance models. %
Pagliari\etal\cite{PagliariCMP21} introduced an approach deciding between local inference and remote offloading, considering execution time and energy consumption, focusing  on Recurrent Neural Networks (RNNs). %
Eshratifar\etal\cite{EshratifarAP21} proposed an energy-aware split computing approach using graph theory, representing DNN structure as a Directed Acyclic Graph (DAG) and solved it as  constrained shortest-path problem to find configurations. %

Our approach differs from these studies by considering a comprehensive set of factors, including total latency, energy consumption, accuracy, optimal split layer, and specific hardware parameters, while also focusing on new NN architectures, such as transformer-based models.

\section{Conclusions}\label{sec:conclusions}
In this paper, we addressed the challenge of deploying ML models on resource-constrained edge devices while optimizing energy consumption and performance. \dynasplit, a framework that leverages split computing and hardware-software co-design, solves the complexity of configuring split layers and hardware parameters by efficiently navigating large configuration spaces. Our two-phase approach, combining offline optimization and online scheduling, was evaluated on a real-world testbed using pre-trained neural networks. The results demonstrated that \dynasplit can reduce energy consumption while still meeting stringent latency requirements. 

Future work could explore cost-sensitive deployment models, such as serverless or containerized cloud services, which introduce cold-start latencies and warm-up overhead. Investigating \dynasplit's performance in dynamic, on-demand environments would provide further insights into resource-constrained applications.

\begin{acks}
Experiments presented in this paper were carried out using the Grid'5000 testbed, supported by a scientific interest group hosted by Inria and including CNRS, RENATER and several Universities as well as other organizations (see \url{https://www.grid5000.fr}).

\end{acks}

\bibliographystyle{ACM-Reference-Format}
\bibliography{references}

\end{document}